# On the Solution of Euclidean Path Integrals with Neural Networks

Gábor Balassa *

*Department of Physics, Yonsei University, Seoul 120-749, Korea*

This paper proposes a numerical method using neural networks to solve the path integral problem in quantum mechanics for arbitrary potentials. The method is based on a “radial basis function” expansion of the interaction term that appears in the Euclidean path integral formalism. By constructing a corresponding “multilayered perceptron”-type neural network with exponential nonlinearities in the hidden layer, the original path integral can be approximated by a linear combination of Gaussian path integrals that can be solved analytically. The method has been tested for the double-well potential that includes a quadratic and a quartic term, giving very good, within a few percent, agreement between the true and estimated bound state wave functions that are extracted from the propagator at large Euclidean times. The proposed method can also be used to describe potentials that have imaginary parts, which is tested for a simple Gaussian path integral with complex frequencies, where the model uncertainty stays below 1% for both the real and imaginary parts of the propagator.

## 1. Introduction

Path integral methods are fundamental theories in describing many physical phenomena where the underlying system is governed by a probabilistic/dynamical evolution [1–3]. Perhaps the most well known application of path integrals is the formulation of quantum mechanics, where the evolution of a quantum system is described by a weighted sum over all possible paths a particle could take between some initial and final states [4]. This description provides an alternative framework to the original operator-based methods where the evolution of the system is described on a Hilbert space through operators acting on wave functions. In quantum field theory, the particle trajectories are extended to field configurations, which provides the basis for perturbative techniques, Feynman rules, and nonperturbative methods like lattice field theory [5]. Apart from quantum theories, path integrals are widely used in other areas as well, e.g. in statistical physics [6], stochastic processes in heat transport [7], finance [8], or in the description of biological systems [9].

In their rigorous mathematical form, path integrals are infinite-dimensional functional integrals that have analytical solutions only in very specific cases, e.g. for the harmonic oscillator, linear potential terms, or for the free particle [10]. Real-life systems, however, often consist of higher-order or more exotic interactions that are usually impossible to solve without relying on perturbative or numerical techniques. Perturbation theory, on one hand, offers a very good and tractable method for weak nonlinearities; however, with each additional term, the com-

*balassa.gabor@yonsei.ac.kr

Funded by SCOAP³

plexity grows exponentially, and after the few lowest-order approximations, the calculation of the path integrals becomes very challenging [11]. On the other hand, state-of-the-art numerical techniques, e.g. lattice methods in quantum chromodynamics, offer a general method that is able to handle essentially any type of interaction by also taking into consideration local and global symmetries; however, it suffers from the fact that it relies on a fine discretization of the corresponding phase space to be able to give estimates with manageable noise levels [12–14]. Because of these numerical issues, the computational times even for the simplest systems that are of interest become very large even with modern supercomputers.

Another huge drawback of lattice methods is that they are relying on the Wick-rotated Euclidean version of path integrals that provides a way to connect quantum mechanics with statistical physics; however, by doing so, it will limit the possible interactions to purely real in order that a probabilistic interpretation of the paths can be given [15]. If for some reason an imaginary part appears in the Euclidean path integral, then numerical problems could appear, e.g. the famous "sign problem" in interacting systems at finite densities [16]. Several methods exist that try to overcome this problem, e.g. stochastic quantization [17,18], reweighting [19], etc., each having some success in specific regions or for some specific theories.

For the reasons mentioned above, it is very desirable to find alternative methods that are able to address at least some of these issues. One way would be to directly address the path integral without relying on the discretization of the paths or field configurations and try to estimate the functional integral for specific boundary conditions. Recently it was proposed that the path integral problem in more than 1 dimension could be approached by finite element techniques [20] used predominantly in solving partial differential equations that arise mostly in engineering applications, e.g. heat transfer, electromagnetic scattering, etc. Another interesting method to estimate the Euclidean propagators in nonrelativistic quantum mechanics is shown in Ref. [21], where a deep neural network architecture is used to generate the dominant paths that are used to estimate the quantum propagators. There are also increasing attempts to use different neural network architectures to describe field-theoretical problems in particle and condensed matter physics [22,23].

One of the main goals of this and hopefully many forthcoming works is to apply seemingly nonrelated mathematical methods in particle and nuclear physics that are often omitted in this field but used in e.g. engineering in modeling complex, nonlinear systems [24,25]. One example would be the application of nonautonomous nonlinear Volterra series with the help of neural networks in the field of inverse quantum scattering [26], where the interaction potential is estimated through modeling a nonlinear differential equation that describes the evolution of the phase function in coordinate space. In Ref. [27] the same nonlinear dynamical identification problem is tackled by radial basis function neural networks in Fourier space, or in Ref. [28] with multilayered perceptron-type neural networks in coordinate space.

In this paper we will address the path integral problem in nonrelativistic quantum mechanics for arbitrary potentials by approximating the interaction terms with neural networks in such a way that the resulting path integral will be analytically solvable. The method does not rely on any discretization; therefore, it omits the standard problems that often arise in those cases. Because of its approximate nature, the method will still have an "operating range" where it can be used because of the fact that the identification can only be done by finite samples in a finite phase space. This, however, will not be a problem for physically sensible potentials, e.g. those that are bounded from below. Because of the very general nature of the neural networks, even this constraint can be overcome by careful modeling of the underlying system.

In Section 2 the physical and mathematical backgrounds of path integrals and neural networks are briefly summarized, then in Section 3 the general model and the corresponding neural network architecture are described. After a brief description of the model, it will be used in Section 4, where in Section 4.1 the neural network model will be applied to the double-well potential case, where the bound state wave functions will be extracted through the estimated Euclidean propagators. In Section 4.2, the complex potential case will also be addressed by estimating the real and imaginary parts of the propagator in the case of the harmonic oscillator with complex frequencies. In Section 5, the approximation capabilities and time complexity of the neural network model is further discussed. Finally, in Section 6, we briefly conclude the work and its future possibilities, especially in quantum field theories.

## 2. Physical and mathematical background

In this section the physical motivation of path integrals, their mathematical background, and the necessary mathematical tools to address them with neural networks will be briefly summarized. As both path integral methods and neural networks are well known and have quite a large literature, here we will only consider the necessary basics that are needed for the understanding of the proposed model that is used to estimate the Euclidean propagator in quantum mechanics for a wide range of potentials.

### 2.1. *Path integrals in quantum mechanics*

The path integral formulation in quantum mechanics is an alternative approach to the usual operator-based methods, where instead of calculating wave functions and energy eigenvalues by solving the Schrödinger equation, the fundamental quantity that is of interest is the quantum propagator that describes the probability amplitude to find a particle in a specific final state $|x_f\rangle$ after some $T = t_f - t_i$ time when starting from some initial state $|x_i\rangle$. In this formalism the propagator depends on all of the possible paths that a particle could take between its initial and final states and can be formulated in the following functional integral:

$$K(x_f, t_f, x_i, t_i) = \int\limits_{\substack{x(t_f)=x_f \\ x(t_i)=x_i}} \mathcal{D}x(t) e^{\frac{i}{\hbar} S(x(t))}, \tag{1}$$

where the $\int \mathcal{D}x(t)$ integral represents the sum over all possible $x(t)$ paths that are allowed by local/global symmetries and satisfy the given boundary conditions. The measure in this form is not a well defined object but can be defined more precisely in the limit of large $N$ after discretizing the paths into $N + 1$ time slices:

$$\int \mathcal{D}x(t) = \lim_{N\to\infty} \left(\frac{m}{2\pi i\hbar\Delta t}\right)^{\frac{N+1}{2}} \int_{\mathbb{R}^N} \prod_{k=1}^{N} dx_k, \tag{2}$$

where the full path is separated into $N + 1$ subpaths that follow each other as the time moves forward $t_1, t_2, \ldots, t_k, \ldots$ with $dx_k = dx(t_k)$. The physical meaning of such an object is that every possible path corresponds to the total probability with a weight factor that is related to the classical action $S(x, t)$, where the action is given by the Lagrangian as:

$$S(x(t)) = \int_{t_i}^{t_f} dt\ L\Big(x(t), \frac{dx(t)}{dt}\Big) \tag{3}$$

where $L(x, dx/dt) = T_K - V$ is the classical Lagrangian (where $T_K$ is a kinetic term, while $V$ is the potential term) that describes the dynamics through the time derivatives of $x(t)$ and the interactions through some potential function. The propagator in this form describes the time evolution of the dynamical system in coordinate space, which means that if one knows an initial wave function at time $t_i$, then the final state wave function at $t_f$ can be calculated by the following integral:

$$\psi(x, t_f) = \int dx' \, K(x_f, t_f, x', t_i)\psi(x', t_i), \tag{4}$$

where $\psi(x, t)$ represents the wave function that also appears in the time-dependent Schrödinger equation [29]. The propagator also has to respect the symmetries of the system; e.g. for potentials that are purely real-valued, it has to respect unitary evolution; however, for systems that could consist of dissipative terms, e.g. through an imaginary part of the potential, this is no longer satisfied, and the propagator will not preserve the norm of the wave function during its evolution [30,31].

The path integral in this form is directly related to the Schrödinger equation; however, because of the heavily oscillating integrand, it is very hard to work with from a numerical standpoint. To overcome this problem, the Wick rotation is introduced, where the real-time coordinate is replaced by an imaginary time through the transformation $t \to -i\tau$, which will change the original action to the Euclidean action as $S_E = iS$, in which case the path integral becomes:

$$K_E(x_f, x_i, T) = \int\limits_{\substack{x(T)=x_f \\ x(0)=x_i}} \mathcal{D}x(\tau) e^{-\frac{1}{\hbar} S_E(x(\tau))}, \tag{5}$$

where, without loss of generality, the initial and final Euclidean times are set to 0 and $T$, and $K_E$ represents the Euclidean propagator that now does not contain the oscillatory term, but instead depends on the exponentially damped $e^{-S_E/\hbar}$ weight factors. The Euclidean action now has the form of $\int d\tau(T_K + V)$ instead of $\int d\tau(T_K - V)$. These changes will prove very useful when one wants to calculate path integrals on a lattice through Monte Carlo sampling [32]; however, the problem of summing infinitely many paths still remains. This can be partially overcome, as the form of the path integral indicates that the most important paths will be the classical ones and the ones with small fluctuations around them; however, in interesting systems in more than 1+1 dimensions, the time to do lattice calculations is still severe.

The Euclidean version of the path integral formalism is not directly related to the Schrödinger equation; however, by introducing the inverse temperature as $\beta = \frac{\hbar}{k_B T_H}$, where $k_B$ is the Boltzmann constant and $T_H$ is the temperature, the quantum partition function can be expressed as:

$$Z(\beta) = Tr\left(e^{-\beta H}\right) = \oint\limits_{x(0)=x(\hbar\beta)} dx \, K_E(x, x, \hbar\beta), \tag{6}$$

where the integration now means that we have to sum all the possible periodic paths with a period $\hbar\beta$. By knowing the partition function, the thermodynamics of the original system can be studied through the derivatives of the partition function [33]. Apart from the thermodynamic quantities, the bound state energy of the system is also accessible through the spectral decomposition of the partition function and by taking the large-$\beta \to \infty$ limit in the free energy as:

$$\lim_{\beta\to\infty} F(\beta) = -\lim_{\beta\to\infty} \frac{1}{\beta} \log Z(\beta) = -\lim_{\beta\to\infty} \frac{1}{\beta} \log \sum_n e^{-\beta E_n} = E_0, \tag{7}$$

where $F(\beta)$ is the free energy, and $Z(\beta) = \sum_n e^{-\beta E_n}$ is the spectral decomposition of the partition function. Similarly, by expanding the propagator in terms of the eigenstates of a given Hamiltonian, the spectral decomposition of the propagator can be given as follows:

$$K_E(x_f, x_i, T) = \sum_n \psi_n(x_f)\psi_n(x_i)^* e^{-E_n T}, \quad (8)$$

where $T$ is the Euclidean time, $\psi_n(x)$ is the wave function of the $n$th bound state, and $E_n$ is the corresponding energy eigenvalue, and the sum consists of all the bound and scattering states as well. In the large time limit, when $T \gg \hbar/\Delta E$, where $\Delta E$ is the energy gap between the bound state and the first excited state, the propagator can be related to the bound state wave function as:

$$|\psi_0(x)|^2 \propto K_E(x, x, T), \quad (9)$$

where $\psi_0(x)$ is the ground state wave function, and $T$ is some sufficiently large Euclidean time. Taking into consideration the large time behavior of the spectral decomposition, the bound state energy and the bound state wave functions are both pieces of information that can be extracted by solving the corresponding path integral.

In the case of a dissipative or certain kind of open system, such as a complex scalar field at finite density, the interactions can introduce imaginary terms into the Euclidean path integral formalism, making numerical solutions more involved. The model that is proposed here is able to tackle both types of systems and does not make a distinction between complex or purely real potentials, which will be shown in Section 3.2. It will also be shown that the method is very general and there is a possibility to extend its scope to be able to describe path integrals and correlation functions in quantum field theory as well, which of course generates a new set of problems, and is left for future work.

### 2.2. *Neural networks in physics*

With the increasing numerical capabilities in the past few decades, machine learning techniques have become widely used methods in modeling complex systems where the underlying system is only fairly well understood or fast numerical methods are needed with good generalization capabilities [34–36]. Many real-life systems in physics can be described by linear or nonlinear dynamical models that are governed by their corresponding differential equations, which can be modeled in many different ways, e.g. state-space models [37], autoregressive models [38], nonlinear Volterra models [39], neural network models, etc. Neural networks are part of many machine learning techniques that, in their original form, try to mimic how the human brain processes gathered information as an input–output system [40]. Nowadays, there are numerous neural network architectures available, e.g. multilayered perceptron-type feedforward networks (MLP) [41], radial basis function networks (RBF) [42], convolutional networks (CNN) [43], recurrent networks (RNN) [44], and so on. Recently, physics-informed neural networks are also becoming very popular because of their ability to incorporate physics-based constraints into the training procedure [45]. Each of these constructions has its advantages and drawbacks and, in practice, each problem requires a careful analysis of which model could be used effectively.

In this paper a specific type of MLP network is used to tackle the path integral problem, which is chosen because of its specific form that could be applied to describe the appearing functional form that we want to model. In their mathematical form, MLP networks consist of multiple interconnected layers, each having a specific number of so-called neurons that introduce non-linearity into the system through their activation functions. The connections between layers and their neurons are each given a weight factor (and a bias) that tells us how important the corresponding neuron is in the final response of the system. There are several types of activation functions that can be used, e.g. sigmoid, rectified linear unit (ReLU), exponential, polynomial, etc., each having advantages and disadvantages that have to be considered during the modeling step. The general topology of a feedforward network for a multiple input–multiple output system with $k$ layers is shown in Fig. 1, where the inputs are labeled as $x_1, x_2, \ldots, x_n$ and the outputs are $y_1, y_2, \ldots, y_m$. The first layer that contains only inputs is called the input layer, and the last layer is called the output layer, while all the in-between $(k-1)$ layers are generally called hidden layers.

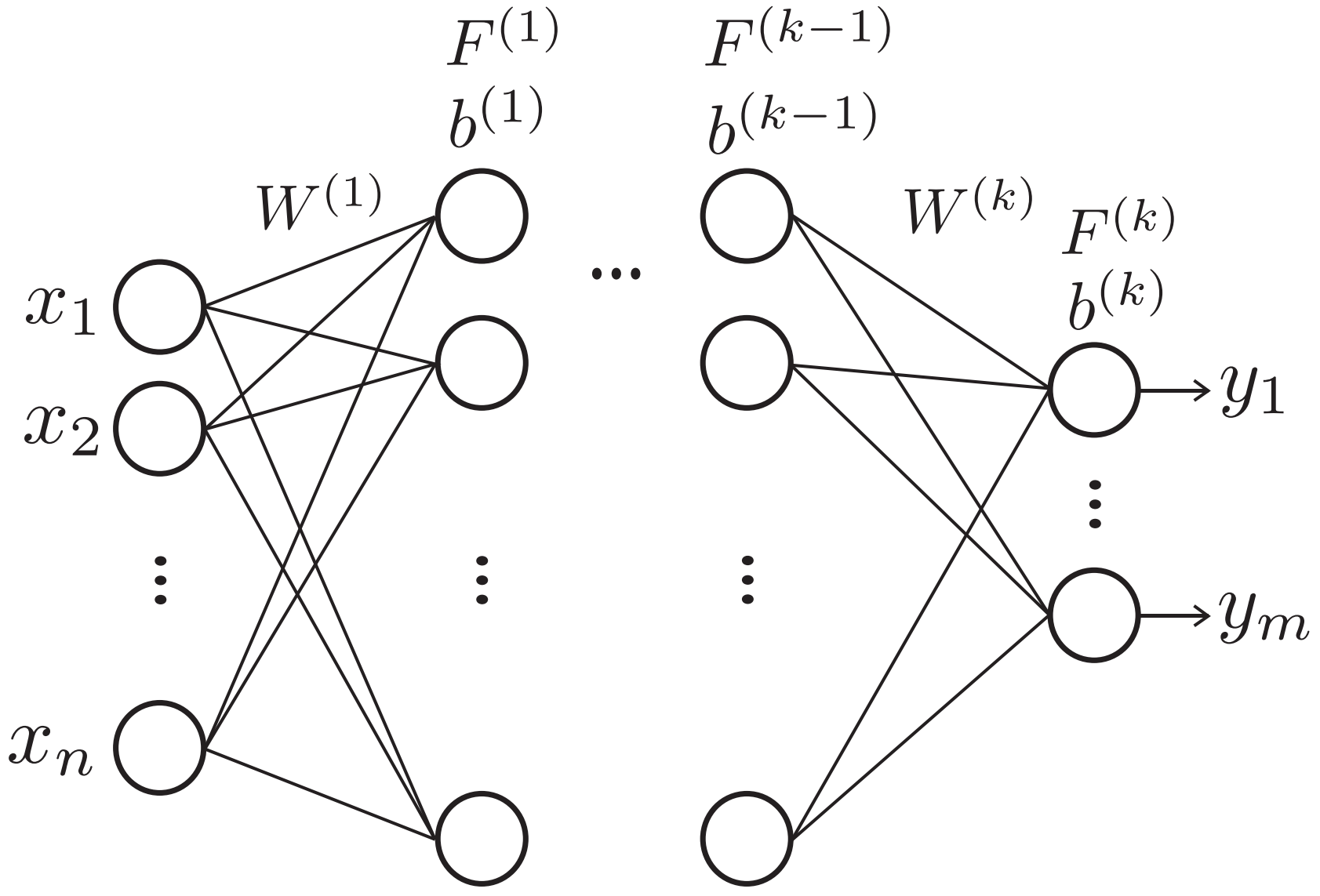

**Fig. 1.** Schematic view of a feedforward multilayered perceptron-type neural network with $n$ inputs, $m$ outputs, and $(k-1)$ hidden layers.

Figure 1 shows a feedforward MLP network with $k$ layers, $n$ inputs, and $m$ outputs, where each hidden layer could have different number of $n_k$ neurons. The general mathematical form of this architecture can be given as:

$$\mathbf{y} = \mathbf{F}^{(k)}(\mathbf{b}^k + \mathbf{W}^k\mathbf{F}^{k-1}(\mathbf{b}^{k-1} + \mathbf{W}^{k-1}\mathbf{F}^{k-2}(\ldots\ldots\mathbf{F}^1(\mathbf{b}^1 + \mathbf{w}^1\mathbf{x})))), \tag{10}$$

where $\mathbf{F}^{(k)} = (f_1^{(k)}, \ldots, f_{n_k}^{(k)})$ is a vector that contains $n_k$ activation functions in the $k$th layer, $\mathbf{b}^{(k)} = (b_1^{(k)}, \ldots, b_{n_k}^{(k)})$ is a bias vector of the $k$th layer, and $\mathbf{W}^{(k)}$ is a weight matrix that is constructed from the $w_{ij}^{(k)}$ weights in each layer, while $\mathbf{x} = (x_1, x_2, \ldots, x_n)$ is the input and $\mathbf{y} = (y_1, y_2, \ldots, y_m)$ is the output vector. The output layer is usually taken to be a linear combination of the outputs from the last hidden layer; however, it can be easily adjusted to specific constraints; e.g. if the output should be strictly positive, then a positive, differentiable function such as the $\log(1 + e^x)$ softplus function could be used instead of the linear activation function. Certain applications could require certain types of activation functions in the hidden layers as well, and it requires careful analysis and testing to find the optimal ones that are adequate to the problem at hand. The feedforward MLP structure shown here has been proven to be a universal approximator [46] that is able to approximate any complex functional dependences between the inputs and outputs; however, the proof does not tell us the complexity of the applicable network, e.g. the number of hidden layers or the number of neurons. Interestingly, it can also be shown that certain types of neural networks are also equivalent to a Volterra series representation of the nonlinear dynamical system [47] by Taylor expanding the activation functions around their bias values, which can be used to extract the higher-order Volterra kernels by identifying a neural model, thus giving us another point of view on how neural networks could be used to approximate any nonlinear input–output relationship.

Training the neural network means that one has to find an optimal parameter set (weights and biases) in a sense that minimizes some predefined error function, e.g. absolute, relative, mean squared errors, etc. There are several types of training methods that can be used depending on the problem at hand, e.g. supervised learning, unsupervised learning, reinforcement learning, or hybrid methods [48]. Here, we will consider supervised learning [49], which means that labeled input–output data pairs are used in the training procedure, where in each iteration the estimate of the true output is calculated, and then according to the value of the loss function, a new parameter set is calculated in each iteration as long as a global or local minimum is achieved or some stopping condition is reached. One of the most used optimization techniques for these kinds of problems is the gradient-based method [50], e.g. gradient descent, stochastic gradient descent, conjugated gradient descent, adaptive gradient descent, etc. All of these methods use the gradient of the loss function to adjust the model parameters by using e.g. the backpropagation algorithm. One of the main advantages of using neural networks is their ability to generalize to unknown input data. To achieve good generalization, many techniques could be used, e.g. batch normalization, dropout methods, or L2-regularization [51], etc.; however, it has to be noted that if a network's architecture is too simple, it can underfit to the data or, in contrast, if it is too complex, e.g. it has too many neurons and/or hidden layers, the network could overfit. To overcome this problem, in addition to regularization techniques, the data are usually distributed into training, validation, and test sets.

The specific network architecture and training procedure will be described in detail in Section 3, where it will be shown that a certain type of feedforward MLP network with exponential activation functions and carefully constructed inputs can be transformed into a corresponding radial basis function (RBF) network that can be used to estimate the interaction part of the path integral problem.

## 3. Neural network expansion of the path integral

In this section a feedforward MLP architecture is constructed to estimate the infinite sum of weighted paths that appears in the path integral formalism in quantum mechanics. The general approach will be described in Section 3.1; then in Section 3.2, a detailed description of the training procedure, e.g. the generation of training samples, the validity of the approximations, etc., will be given. From this point on in the description of the physical systems, the natural units ($\hbar = c = 1$) convention will be used, and we drop the $E$ subscript from the Euclidean propagator $K_E$ and Euclidean action $S_E$ terms for simplicity, as we will only consider Euclidean path integrals from now on.

### 3.1. *General method*

The Euclidean path integral for a quantum mechanical system in general can be written in the following form:

$$K(x_f, x_i, T) = \int\limits_{\substack{x(T)=x_f \\ x(0)=x_i}} \mathcal{D}x(\tau)\exp\left(-\int_0^T d\tau\left(\frac{m}{2}\dot{x}^2 + V(x)\right)\right), \tag{11}$$

where $T$ is the Euclidean time, $x_f$ and $x_i$ are the final and initial positions given by the boundary conditions, $m$ is the mass of the particle, $\dot{x}$ represents the time derivative, while $V(x)$ is a general potential term that usually depends on the coordinate of the particle and can be real, imaginary, or general complex as well. In the exponent, the Euclidean time dependence of the paths $x = x(\tau)$ is understood. The starting point of the model is the knowledge that the path integral problem can be solved analytically for specific potentials, e.g. linear $V(x) = ax$ or quadratic $V(x) = ax^2 + bx + c$, or in the free particle case, when the potential is zero everywhere. In the general case, the potential is some nonlinear function that makes the corresponding equation of motion a nonlinear differential equation; there is usually no analytical solution available; therefore, numerical techniques are necessary to estimate the propagators.

The idea is that if one could estimate the functional form that is inside the path integral by some expansion or a set of basis functions that has a form for which analytical integration is possible, then the path integral can be done and the propagator could be estimated in a closed form. To make this more explicit, let us write the exponential term inside of the path integral in the following form, where the interaction term is separated from the kinetic term:

$$\exp\left(-\int_0^T d\tau\left(\frac{m}{2}\dot{x}^2 + V(x)\right)\right) = \exp\left(-\int_0^T d\tau\,\frac{m}{2}\dot{x}^2\right)\cdot\exp\left(-\int_0^T dt\,V(x)\right), \tag{12}$$

where now the kinetic term is separated from the interaction term that is the exponential of the Euclidean time integral of the potential term. Next, let us approximate this interaction term by a sum of Gaussian basis functions as:

$$\exp\left(-\int_0^T d\tau\,V(x)\right) \approx \sum_{k=1}^{M} a_k\exp\left(-\int_0^T d\tau\,b_k(x-c_k)^2\right), \tag{13}$$

where $a_k$, $b_k$, and $c_k$ are free parameters, where $a_k$ controls the weight of how much a specific exponential contributes to the final result, $b_k$ controls the width or operating range of a specific term, while $c_k$ is the center of that term, where it is mostly active. This form resembles a radial basis function neural network architecture with Gaussian basis functions; however, the integral in the exponential terms makes things a bit less trivial. The approximating capabilities of such models will be discussed in more detail in Section 5.

By putting back the radial basis expansion into the full path integral in Eq. (11), we will arrive at the following weighted sum of quadratic path integrals:

$$K(x_f, x_i, T) \approx \sum_{k=1}^{M} a_k \int\limits_{\substack{x(T)=x_f \\ x(0)=x_i}} \mathcal{D}x(\tau)\exp\left(-\int_0^T dt\left(\frac{m}{2}\dot{x}^2 + b_k(x-c_k)^2\right)\right), \tag{14}$$

where we have used the fact that the sum and the linear weight $a_k$ can be taken outside of the path integral. The path integral now is reorganized into $M$ shifted Gaussian path integrals that can be solved analytically by expanding the paths to a classical one $x_c(\tau)$ and to quantum fluctuations $y(\tau)$ around the classical trajectories, as $x(\tau) = x_c(\tau) + y(\tau)$. In the case of quadratic Lagrangians where the property $\frac{\delta^{(n)}S}{\delta x^n} = 0$ is satisfied for $n > 2$, the path integral could be written in the following simplified form [52]:

$$Q(x_f, x_i, T) = e^{-S(x_{cl})} \int \mathcal{D}y\, e^{-\frac{1}{2}\frac{\delta^2 S}{\delta x^2}y^2}, \tag{15}$$

where $S(x_{cl})$ is the Euclidean action evaluated on the classical path, and $Q(x_f, x_i, T)$ represents the propagator for the path integral with quadratic Lagrangians. The first part only depends on the classical action and the boundary conditions, while the second term is a path integral of the quantum fluctuations that does not depend on $x_i$ and $x_f$ anymore, as it only measures the fluctuations with fixed $x_i = 0$, and $x_f = 0$ endpoints. The solution of the first term is quite straightforward for the shifted harmonic oscillator. First we have to determine the classical equation of motions for the corresponding Euclidean Lagrangian by using the Euler–Lagrange equations; then we have to solve the related differential equations with the given boundary conditions.

To illustrate this, let us take the Euclidean Lagrangian that corresponds to one quadratic term in Eq. (14), e.g. $L = \frac{1}{2}m\dot{x}_{cl}^2 + \frac{1}{2}m\omega^2(x_{cl} - s)^2$, where $x_{cl}$ represents the classical path. Comparing this form to Eq. (14), it can be seen that the frequency corresponds to $\omega = \sqrt{\frac{2b}{m}}$, while the shift is simply given by $s = c$, where we have dropped the $k$-indices for simplicity. In this case, by using the Euler–Lagrange equations, the equations of motion become:

$$\ddot{x}_{cl} = \omega^2(x_{cl} - s), \tag{16}$$

where $\omega$ is the frequency of the oscillating system in the original Minkowskian space, while $s$ is a shift parameter related to the equilibrium point. The solution for this second-order differential equation with the boundary conditions $x_{cl}(0) = x_i$ and $x_{cl}(T) = x_f$ can be given as follows:

$$x_{cl}(\tau) = (x_i - s)\cosh(\omega\tau) + \frac{x_f - s - (x_i - s)\cosh(\omega T)}{\sinh(\omega T)}\sinh(\omega\tau) + s, \tag{17}$$

where the solution is now a linear combination of sinh and cosh hyperbolic functions, which is the consequence of the Euclidean formalism. The general shape of $x_{cl}(\tau)$ will be a so-called "chain" curve that, apart from the $b$ and $s$ parameters, will depend on the $m$ mass of the particle through the $\omega$ "frequency" term. By taking the time derivative of $x_{cl}(\tau)$, then collecting the different terms, the Euclidean action can be expressed in the following closed form:

$$\begin{aligned} S(x_{cl}) &= \int_0^T d\tau \left(\frac{1}{2}m\dot{x}_{cl}^2 + \frac{1}{2}m\omega^2(x_{cl} - s)^2\right) = \\ &= \frac{m\omega}{\sinh(\omega T)}\left[\cosh(\omega T)\left(s^2 + \frac{x_f^2 + x_i^2}{2} - s(x_f + x_i)\right) + s(x_f + x_i) - s^2 - x_f x_i\right] \end{aligned} \tag{18}$$

where we have assumed that $m > 0$ and $\omega > 0$. The Euclidean action shown here will represent the action that is evaluated on the classical path $x_c$; therefore, the next step is to determine the fluctuations around $x_c$. The second term, including the quantum fluctuations around the classical path in Eq. (15), is not that straightforward to calculate because of the functional determinant $\left(\det \frac{\delta^2 S}{\delta x^2}\right)^{-1/2}$ that arises after doing the Gaussian integral. The solution, however, can still be given in a closed form as:

$$\int \mathcal{D}y\, e^{-\frac{1}{2}\frac{\delta^2 S}{\delta x^2}y^2} = \sqrt{\frac{m\omega}{2\pi\sinh(\omega T)}}, \tag{19}$$

which does not depend on the boundary conditions or on the shift parameter. Putting together the classical part and the quantum fluctuations, the path integral for the shifted quadratic potential could be given as:

$$Q(x_f, x_i, T) = \sqrt{\frac{m\omega}{2\pi \sinh(\omega T)}} \exp\left( -\frac{m\omega}{\sinh(\omega T)} \times \right.$$
$$\left. \times \left[ \cosh(\omega T)\left(s^2 + \frac{x_f^2 + x_i^2}{2} - s(x_f + x_i)\right) + s(x_f + x_i) - s^2 - x_f x_i \right] \right). \quad (20)$$

Using this result, the full path integral in Eq. (11) that is approximated by the linear combination of $M$ quadratic path integrals in Eq. (14) can be expressed in the following closed form:

$$K(x_f, x_i, T) \approx \sum_{k=1}^{M} a_k Q_k(x_f, x_i, T) = \sum_{k=1}^{M} a_k \sqrt{\frac{m\omega_k}{2\pi \sinh(\omega_k T)}} \exp\left( -\frac{m\omega_k}{\sinh(\omega_k T)} \times \right.$$
$$\left. \times \left[ \cosh(\omega_k T)\left(s_k^2 + \frac{x_f^2 + x_i^2}{2} - s_k(x_f + x_i)\right) + s_k(x_f + x_i) - s_k^2 - x_f x_i \right] \right), \quad (21)$$

where the $k$-indices in the $Q_k(x_f, x_i, T))$ functions mean that we have different $\omega_k = \sqrt{\frac{2b_k}{m}}$ frequencies and different $s_k = c_k$ shifts for each kernel. These results for the radial basis expansion and the general closed-form result for the path integral for the shifted harmonic oscillator potential will be the basis for the neural network expansion that will be described next.

Starting from the radial basis function expansion in Eqs. (13) and (14), let us construct a feedforward MLP network that will give us a more general expansion, but with the same possibility of expressing the output as a linear combination of quadratic path integrals, so the result in Eq. (21) could be immediately applied to the more general case that will follow. The general topology of the proposed neural network is shown in Fig. 2, where we have introduced $N$ inputs in the form of $u_i = \int_0^T d\tau\, (x - c_i)^2$, each shifted by a $c_i$ parameter, then squared and integrated out from 0 to $T$ in Euclidean time $\tau$.

The proposed network is a fully connected, one-hidden-layered MLP with $N$ inputs, each connected to all of the neurons in the hidden layer through an associated weight parameter $b_{ik}$, where $i$ represents the $i$th input, while $k$ represents the $k$th neuron in the hidden layer. Each neuron in the hidden layer is then connected to the output layer with associated weights $a_k$. Additionally, each neuron in the hidden and output layers consists of a bias term represented by $d_0$ and $d_k$. The activation functions in the hidden layer are chosen to be $f_i(x) = \exp(-x)$ exponential functions, while the output layer is set to be a linear combination of the outputs of the hidden layer. The output of the proposed neural network architecture can be given as follows:

$$y = d_0 + \sum_{k=1}^{M} a_k \exp\left( -d_k - \sum_{i=1}^{N} \left[ b_{ik} \int_0^T d\tau\, (x - c_i)^2 \right] \right), \quad (22)$$

where now we have a more general but still quadratic formula for the outputs. To get rid of the summation of the shifted and squared integral terms in the exponential and give a more

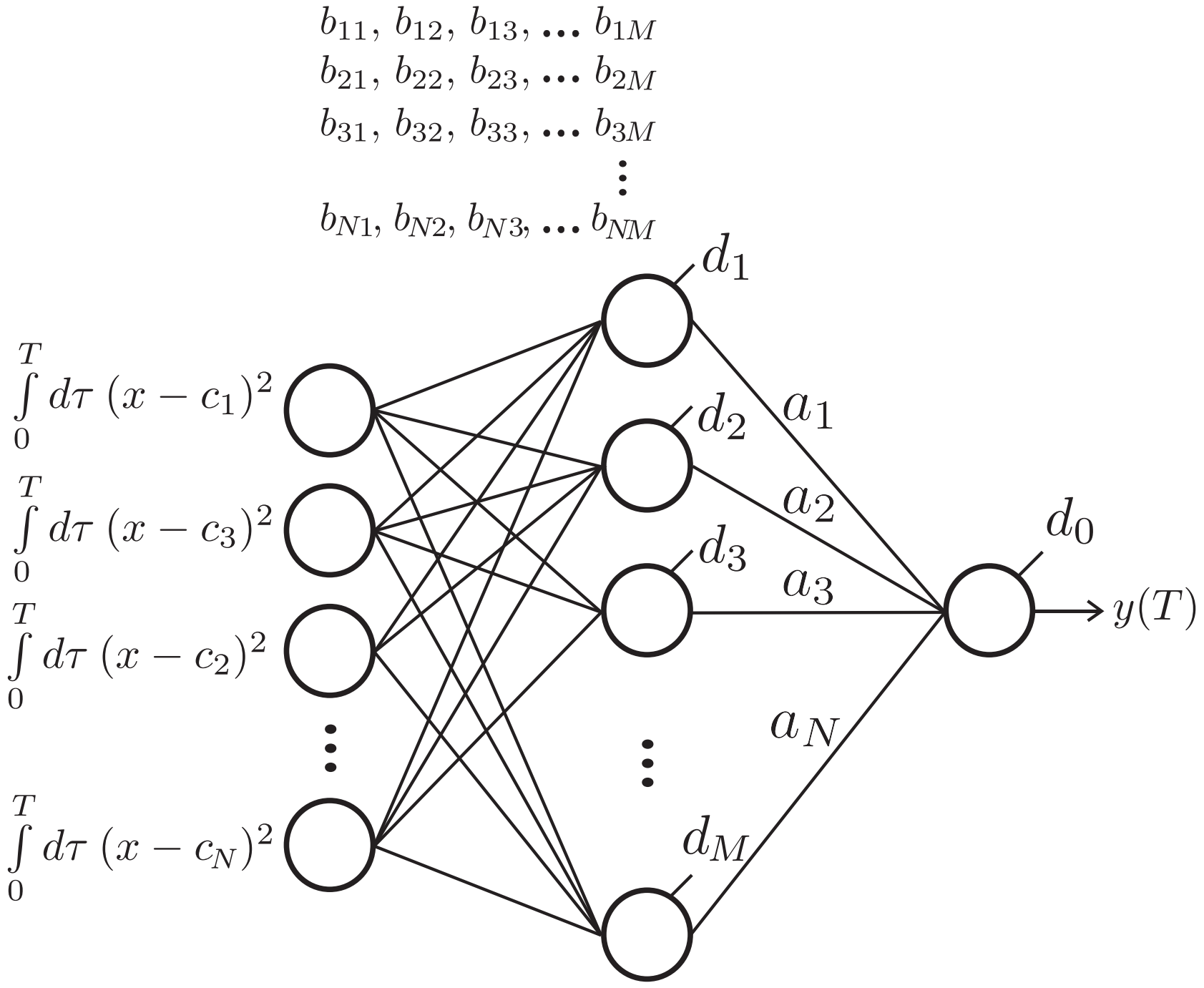

**Fig. 2.** The topology of the proposed neural network with $n$ inputs, 1 output, and one hidden layer with $f_i(x) = e^{-x}$ nonlinearity at each neuron in the hidden layer, and a linear activation function for the only neuron in the output layer. The $b_{ij}$ parameters represent the weight matrix, while the $d_i$ parameters are the biases acting on each neuron.

compact, general quadratic form, let us expand it as follows:

$$\sum_{i=1}^{N}\left[b_{ik}\int_0^T d\tau\,(x-c_i)^2\right] = \sum_{i=1}^{N}\int_0^T d\tau\left[b_{ik}x^2 - 2b_{ik}c_ix + b_{ik}c_i^2\right] = \int_0^T d\tau\left[A_kx^2 + B_kx + C_k\right], \tag{23}$$

where $A_k = \sum_{i=1}^N b_{ik}$, $B_k = -2\sum_{i=1}^N b_{ik}c_i$, and $C_k = \sum_{i=1}^N b_{ik}c_i^2$, new parameters, are introduced. To go further, let us complete the square of this general quadratic expression and integrate out the parts that are independent of $\tau$ as:

$$\int_0^T d\tau\left[A_kx^2 + B_kx + C_k\right] = \int_0^T d\tau\left[A_k\Big(x + \frac{B_k}{2A_k}\Big)^2 - \frac{B_k^2}{4A_k} + C_k\right] =$$
$$= \left[C_k - \frac{B_k^2}{4A_k}\right]T + \int_0^T d\tau\left[A_k\Big(x + \frac{B_k}{2A_k}\Big)^2\right], \tag{24}$$

where we have rearranged the original sum of quadratic integrals into a $T$-dependent shift and a shifted quadratic integral for each $k$-kernel, where the $A_k$, $B_k$, and $C_k$ parameters depend on the weights and biases of the hidden layer. Putting back this expression into Eq. (22), the response of the neural network can be given as:

$$y = d_0 + \sum_{k=1}^{M} a_k \exp\Big(-d_k - \Big[C_k - \frac{B_k^2}{4A_k}\Big]T\Big)\exp\Big(-\int_0^T d\tau\left[A_k\Big(x + \frac{B_k}{2A_k}\Big)^2\right]\Big), \tag{25}$$

where we have separated the terms that are independent of the paths (first exponential) from the ones that depend on them and cannot be taken out from the path integral. Next, let us use this expansion from the output of the neural network to estimate the interaction term shown in Eq. (13) that appears in the path integral after separating it from the kinetic term.

$$\exp\Big(-\int_0^T dt\, V(x)\Big) \approx d_0 + \sum_{k=1}^{M} \hat{a}_k(T) \exp\Big(-\int_0^T d\tau\, \Big[A_k\Big(x + \frac{B_k}{2A_k}\Big)^2\Big]\Big), \tag{26}$$

where we have introduced the Euclidean time-dependent coefficients $a_k(T)$ as:

$$\hat{a}_k(T) = a_k \exp\Big(-d_k - \Big[C_k - \frac{B_k^2}{4A_k}\Big]T\Big). \tag{27}$$

The expression shown in Eq. (25) is still quadratic but has a more general form than the simple radial basis function expansion that was shown in Eq. (13).

Putting everything together, the path integral for a general $V(x)$ potential shown in Eq. (11) can be approximated by the following expansion given by the feedforward MLP neural network model:

$$K(x_f, x_i, T) \approx d_0 \int_{\substack{x(T)=x_f \\ x(0)=x_i}} \mathcal{D}x(\tau) \exp\Big(-\int_0^T dt\, \Big(\frac{m}{2}\dot{x}^2\Big)\Big)$$
$$+ \sum_{k=1}^{M} \hat{a}_k(T) \int_{\substack{x(T)=x_f \\ x(0)=x_i}} \mathcal{D}x(\tau) \exp\Big(-\int_0^T d\tau\, \Big[\frac{m}{2}\dot{x}^2 + A_k\Big(x + \frac{B_k}{2A_k}\Big)^2\Big]\Big), \tag{28}$$

where $A_k = \sum_{i=1}^{N} b_{ik}$, $B_k = -2\sum_{i=1}^{N} b_{ik}c_i$, and $C_k = \sum_{i=1}^{N} b_{ik}c_i^2$ are constructed from the parameters of the neural network, where $N$ is the number of inputs and $M$ is the number of neurons in the hidden layer, while the $\hat{a}_k(T)$ coefficients are given by Eq. (27). The path integral in this case is the sum of a free propagator with coefficient $d_0$, and a sum of $M$ shifted harmonic oscillators with frequencies $\omega_k = \sqrt{2A_k/m}$, which can be summarized as:

$$K(x_f, x_i, T) \approx d_0 F(x_f, x_i, T) + \sum_{k=1}^{M} \hat{a}_k(T) Q_k(x_f, x_i, T), \tag{29}$$

where $Q_k(x_f, x_i, T)$ is the solution for the shifted harmonic oscillator path integral with mass $m$, frequency $\omega_k = \sqrt{2A_k/m}$, and shift $s_k = -\frac{B_k}{2A_k}$ that has the solution shown in Eq. (20), while $F(x_f, x_i, T)$ is the path integral for the free particle case that has the analytic form of:

$$F(x_f, x_i, T) = \sqrt{\frac{m}{2\pi T}} \exp\Big(-\frac{m(x_f - x_i)}{2T}\Big). \tag{30}$$

To summarize the general method, the feedforward MLP neural network shown in Fig. 2 has to be trained so that its output given by Eq. (25) will be a good estimate of the exponentialized potential in Eq. (26). In the case in which a good estimation is possible, the resulting path integral will be given by the linear combination of a free particle and $M$ shifted quadratic path integrals shown in Eq. (28) with coefficients that depend on the Euclidean time as well. The given path integral therefore has an analytical solution, which will be an approximation of the original path integral problem with a general $V(x)$ potential. Our main task is now to give a prescription on how to achieve a good approximation of this system or, in other words, how to train the neural network so that it has good generalization capabilities and is able to estimate a problem where, in theory, infinite paths should be considered. The next section will provide a method to this exact problem, where it will be shown that by careful construction of a finite number of paths, the network could be trained so that the error for very large (possibly infinite) sums using different paths will be well controlled.

### 3.2. *Training the network*

Training a neural network in practice means that we have to solve an optimization problem for the free parameters (weights and biases) of the corresponding network in the sense of some predefined loss function, e.g. mean squared, absolute, cross-entropy, etc., that is able to describe the input–output relationship in a meaningful way. To do this, one needs a large amount of good-quality training samples that are used to generate the estimated outputs of the neural network, and then, by comparing the estimated outputs to the true outputs, the free parameters can be adjusted so that a global or a local minimum is achieved in the loss function. The training samples have to be able to excite the underlying physical system in a way that the model will be able to give a good generalization to unknown inputs that were not used in the training process as well. This procedure resembles the well known problem in system identification [53], where, e.g. in frequency domain identification, one needs good excitation signals that are able to excite meaningful frequency components that are relevant in practical applications.

To train the feedforward MLP network that is shown in Fig. 2, first, let us summarize the mathematical problem that needs to be solved. The estimated output $y_{\mathrm{est},i}(T)$ that is given by the neural network model for an input sample $u_i(\tau)$ can be given in its full form with all the free parameters expanded as:

$$y_{\mathrm{est},i}(T;\theta) = d_0 + \sum_{k=1}^{M}\Bigg[a_k \exp\Bigg(-d_k - \Big[\sum_{i=1}^{N} b_{ik}c_i^2 - \frac{(\sum_{i=1}^{N} b_{ik}c_i)^2}{\sum_{i=1}^{N} b_{ik}}\Big]T\Bigg) \times \exp\Bigg(-\int_0^T d\tau \Big[\sum_{i=1}^{N} b_{ik}\Big(u_i(\tau) - \frac{\sum_{i=1}^{N} b_{ik}c_i}{\sum_{i=1}^{N} b_{ik}}\Big)^2\Big]\Bigg)\Bigg], \tag{31}$$

where $T$ is the Euclidean time that we want to evaluate the network, and $\theta = (a_i, b_{ik}, c_i, d_i)$ represents the collection of parameters that need to be optimized for. The number of free parameters can be given as:

$$N_\theta = 1 + N + 2M + M \times N, \tag{32}$$

where $N$ is the number of inputs of the network, while $M$ is the number of neurons in the hidden layer. The corresponding true output of the problem can be given by the following integral of the $u_i(\tau)$ samples:

$$y_{\mathrm{true},i}(T) = \exp\Bigg(-\int_0^T d\tau\, V(u_i(\tau))\Bigg), \tag{33}$$

where $V(u_i(\tau))$ represents the potential function evaluated at the given training sample $u_i(\tau)$. Before we set up the final optimization problem, e.g. determining the loss function or the inputs that will be used to generate $y_{\mathrm{est}}$ and $y_{\mathrm{true}}$ outputs, or the Euclidean time interval and the boundary conditions, let us first examine the neural network in relation to the underlying physical system and its applications.

It is apparent that the model depends on the integral of the input samples, and the output is the function of the Euclidean time $T$, which is the upper limit of the corresponding integrals. It is a modeling step on how to approach this, as one could try to optimize the network for a large time interval $T \in [T_{\min}, T_{\max}]$ or just one specific $T_i$ value as well. Decisions like this often require test trainings and cross-validation on different training and validation sets and are a usual method in machine learning. There are several aspects that have to be considered when we decide on how to choose the time intervals of the training. Firstly, the physical problem implies that we need to know several values of the propagator at large times to be able to extract bound state information. Therefore, we do not really need to train the network for small times; however, by not doing so, we could lose some information on the dynamics of the system. Secondly, by using a large time interval, it is possible that the variation in the outputs will be very large, spanning multiple orders of magnitude, because of the exponential functions in the hidden layer; therefore, it will be hard to train the network, and several numerical problems could arise. On the other hand, using a finite time interval helps achieve a better generalization because of the greater variation in the corresponding inputs, but on that account the complexity of the network might be larger (more neurons in the hidden layer, more input shifts) and/or the time that is necessary for training will be larger. Because of these problems, selecting a good time interval might seem a cumbersome task, but in practice it is very simple and straightforward, as we only need to do a few tests with different time intervals to see if the network is capable of modeling the actual system with good precision or not.

Another important realization is that the output of the network does not explicitly depend on the boundary conditions $x_f$ and $x_i$. This information is rather coded in the training samples, which means that if we want to only calculate the path integral for specific $x_f$ and $x_i$, then we train the network with training samples that satisfy those conditions. This does not restrict the generality of the model, as in the path integral only those paths are included in the summation that satisfy the specific boundary conditions; therefore, even if the network is not capable of generalizing to other boundary conditions, it does not matter because those paths are omitted from the final sum. In the Euclidean path integral, the most important boundary conditions are given by $x = x_f = x_i$ because of the fact that the partition function only considers periodic paths. The $x$ value could be anything; however, in practice for potentials that show "good" behavior, e.g. bounded from below, the relevant part of the propagator can be given in a finite interval $x \in [x_{\min}, x_{\max}]$. Just like in the case with choosing a finite time interval, a finite boundary interval where $x_f$ is not necessarily equal to $x_i$ helps achieve a better generalization even if we only need the corresponding $x = x_f = x_i$ values in the partition function. One important thing to note is that because of the construction of the network, the output can be kept as a continuous function of the Euclidean time and the position variables $y(T, x_f, x_i)$; however, this is not necessary, and we could train it separately on a discrete time and position grid as well.

As the properties of the paths that will be used as training samples are deeply connected to the physical system that we want to describe, it is a very important task to generate samples that can mimic the actual paths that have to be considered in the path integrals with specific boundary conditions. Theoretically, one has to sum up an infinite number of paths that could take every possible value; however, in practice, because of the $e^{-S_E}$ form in the Euclidean path integral (or similarly because of the highly oscillating integrals in the Minkowskian path integral), the dominant paths will be the classical ones plus small fluctuations around them. This idea is frequently used in lattice methods where, because of numerical constraints, it is impos-sible to consider all possible paths in the sum. If the classical paths are unknown or there are many possible solutions to the classical equations of motion, then an estimation of the possible interval where the dominant paths could appear is also possible by examining the exponential weight factors. In practice, the dominant paths will depend on the mass and other parameters that appear in the Lagrangian; therefore, the range of the excitations that we should consider will be different for every system.

Taking into consideration all of these factors, we have the following expectations for the training samples. (1) The boundary values should be adjustable. (2) The range/image of the excitation functions should be adjustable. (3) The paths should be continuous and integrable. (4) The paths should be able to consist of high-frequency fluctuations and low-frequency components as well. There exist several methods that are able to satisfy the given constraints, e.g. random phase multisines [54] that are a linear combination of multiple sine functions at different frequencies in a predefined frequency range with randomized phases, in which case a well controlled spectral excitation could be given. To satisfy the boundary conditions, an additional linear term could be introduced, whose free parameters are fitted so that the signal satisfies the given initial and final states as:

$$u(\tau) = a\tau + b + \sum_{i=1}^{K} A_i \sin(2\pi f_i \tau + \phi_k), \tag{34}$$

where $a$, $b$ are the parameters of the linear term, $f_i$ represents the predefined frequencies, $A_i$ are the amplitudes of the different sine waves, while $\phi_k$ are the randomly distributed phases taken from some predefined distribution. When the boundary conditions are $u(0) = x_i$ and $u(T) = x_f$, the $a$ and $b$ parameters can be given in a closed form as:

$$\begin{aligned} a &= \tfrac{1}{T}\Big(x_f - x_i + \textstyle\sum_{i=1}^{K} A_i \sin(\phi_i) - \sum_{i=1}^{K} A_i \sin(2\pi f_i T + \phi_i)\Big), \\ b &= x_i - \textstyle\sum_{i=1}^{K} A_i \sin(\phi_i). \end{aligned} \tag{35}$$

Multisine excitation like this but without the linear term has been used before in other neural network models that were applied to inverse scattering problems in nuclear physics [27], where it was important to represent any possible bounded potential functions that are allowed by the physical constraints. This form can be easily extended to more than 1 dimension; therefore, it could prove very useful in later stages, where we intend to apply the model to quantum field theoretical problems, where the $\phi(x,t)$ field configurations will take over the role of the $x(t)$ paths that are used in nonrelativistic quantum mechanics.

The method used here will be somewhat different than the mentioned multisine method, but it will still be able to give a very general class of continuous functions by using piecewise cubic Hermite interpolating polynomials (PCHIP) [55] between a specific number of fixed control points that will be given randomly for each training sample. The main disadvantage of such a method is that no closed form could be given, as it depends on the actual values of the control points; however, this should not be a problem, as in the case of 1 dimension, the numerical integration that is needed in the generation of the network inputs could be calculated easily. The steps for generating one training sample will be given as follows:

- Generate one random integer ($K$) from a uniform distribution that will represent the number of control points of the training sample.

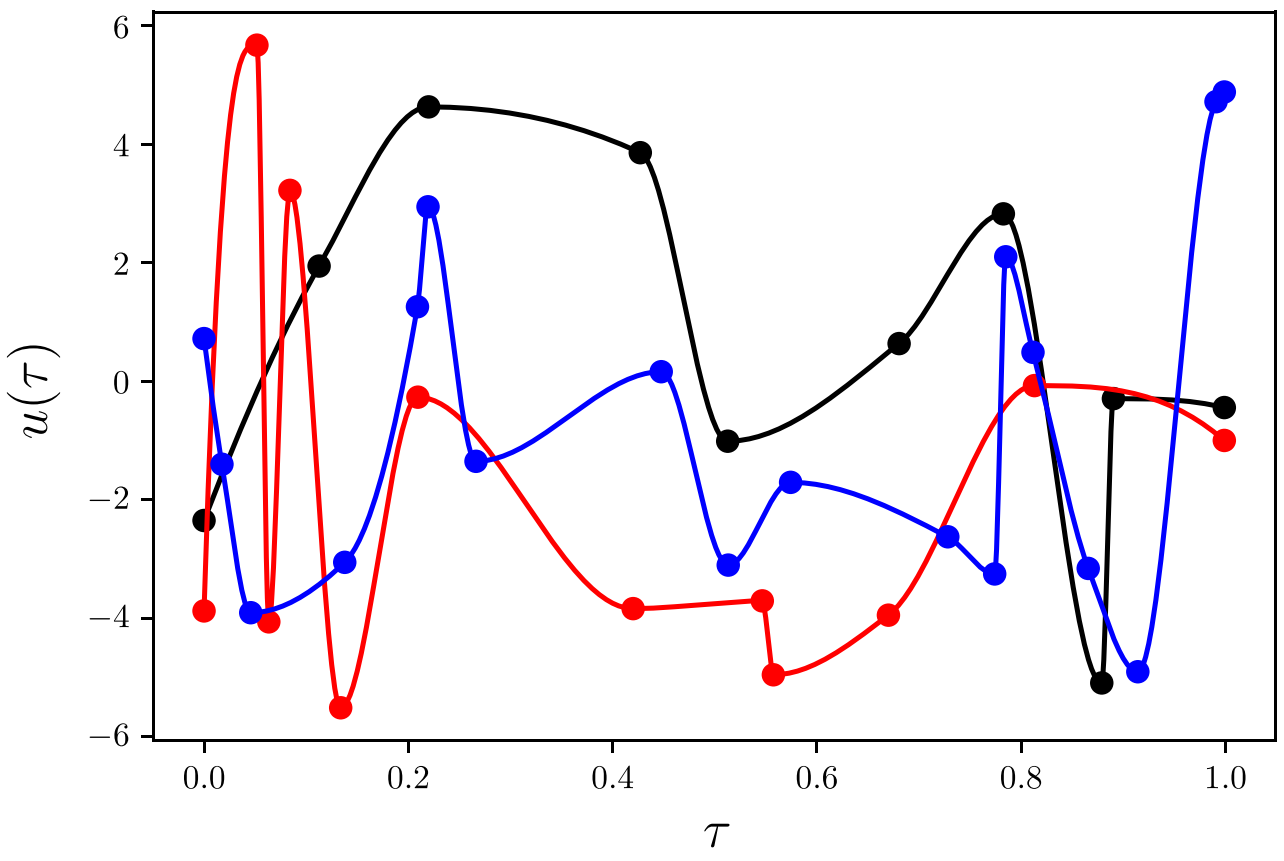

**Fig. 3.** Randomly generated training samples using the PCHIP (piecewise cubic Hermite interpolating polynomial) method.

- Generate $K$ control points in the range of $[0 + \Delta, T - \Delta]$ from a uniform distribution, where $\Delta$ is a small number that is introduced so that we skip the boundary points.
- Set the initial and final points to the boundary conditions $u(0) = x_i$ and $u(T) = x_f$.
- Generate one $u_c(\tau_i)$ value at each control point in the range of $u_c(\tau_i) \in [a, b]$ uniformly.
- Apply a piecewise Hermitian interpolation between the control points, obtaining a continuous $u(\tau)$ training sample, that satisfies the given boundary conditions.

In Fig. 3 three samples are shown that were generated by this method between $\tau \in [0, 1]$ and $u(\tau) \in [-6, 6]$. It can be seen that the generated samples are well randomized; however, their frequency spectrum or its statistical properties are not that straightforward to see, as was the case for the multisine functions. However, there are several advantages to this method. The main advantage is that it is very easy to satisfy any given boundary conditions and function interval bounds because in the PCHIP technique there will be no overshoots between control points that could happen for other interpolators, e.g. splines. This means that we do not have to "renormalize" the obtained functions so that they stay in the desired interval range, which could make the boundaries change as well. If one has to rescale the obtained functions, e.g. because of overshoot values, it could have the disadvantage that the distribution of the initial and final state values will not be uniform, which would have undesirable effects during training and could make generalization harder for the network.

After we have a method to generate the training samples, the next thing is to set up the optimization problem, the constraints (if necessary), and the procedure to solve it. There are several possible constraints that could be taken into consideration given by local or global symmetries, positivity, interval constraints, etc. Each of these constraints could be forced on the network exactly or approximately; however, it has to be kept in mind that it is not always necessary to include every possible constraint in the model because a strict constraint could also hinder the training process, making it slower than in the case when we allow the system to walk through a larger parameter space. Lastly, the constraints are always taken into consideration approximately through a true comparison of the estimated and the outputs. In the basic model that will be used in the next section, only the most necessary constraint will be used, i.e. the positivity of the frequency of all the quadratic path integrals. Indeed, by looking at Eq. (28), we see that the frequency of each quadratic kernel is given by:

$$\omega_k = \sqrt{\frac{2\sum_{i=1}^{N} b_{ik}}{m}}, \tag{36}$$

where $\omega_k$ is the frequency of the $k$th kernel for the corresponding quadratic term with a potential $\frac{1}{2}m\omega_k(x - s_k)^2$, $N$ is the number of inputs of the neural network, $m$ is the mass of the particle, and $b_{ik}$ are the corresponding weights that go from the $i$th input to the $k$th neuron in the hidden layer. To make the frequencies strictly positive, we could make all the corresponding weights positive in the training; however, that would make a severe cut in the possible phase space of the hyperparameters, which, if possible, has to be avoided. A better way to enforce the frequencies to be positive is to simply consider weight combinations, where the corresponding sum will be positive; therefore, the constraint that will be enforced during training can be described as follows:

$$\text{Constraint:} \quad \sum_{i=1}^{N} b_{ik} > \epsilon, \quad \rightarrow \quad \omega_k = \sqrt{\frac{2\sum_{i=1}^{N} b_{ik}}{m}} > 0, \tag{37}$$

where $\epsilon$ is a small positive number that will correspond to the minimum frequency of the harmonic oscillator kernels, and we have assumed that the mass parameter is greater than zero. In the following examples in Section 4, this will be the only constraint that we use to train the model. In the case of any local or global symmetries that would be important to consider in the description of the system, a symmetry-enforcing penalty term could also be added that will force the network to satisfy the given symmetry approximately. If we want to enforce a symmetry exactly, then in some cases, e.g. for rotation or parity transformations, the network could be constructed in a way that the response always satisfies the symmetry properties; however, for more complex transformations, it is not straightforward to construct the network so that it keeps its quadratic nature and still satisfies the symmetries of the system exactly.

Considering all of the above, the optimization problem can be formulated as follows:

$$\theta^* = \underset{\theta}{\operatorname{argmin}} \frac{1}{N_t} \sum_{i=1}^{N_t} E_i\Big[y_{\text{est},i}(T;\theta), y_{\text{true},i}(T)\Big],$$

$$\text{subject to:} \sum_{i=1}^{N} b_{ik} > \epsilon, \tag{38}$$

where $E_i(\cdot)$ is some appropriate error function, $N_t$ is the number of training samples, $N$ is the number of inputs, $\theta$ is the free parameters of the model, $T$ is the Euclidean time, while $y_{\text{est},i}$ represents the output of the neural network given by Eq. (31), and $y_{\text{true},i}$ is the true output of the system that can be calculated from the potentials, as was shown in Eq. (33). Because of the specifics of the numerical problem at hand, the usual mean squared error (MSE) would not be the best choice, as the exponential function could provide data that are imbalanced and could cover a huge interval; therefore, it is possible that there will be large outliers in the training samples as well, which could be problematic from a numerical point of view when we square the differences between the true and estimated outputs. In this case, the absolute difference (L1 loss) is a more robust choice, and in the applications that will follow, the neural network is always trained by using the absolute differences $E_i = |y_{\text{est},i}(T;\theta) - y_{\text{true},i}(T)|$.

One final remark is in order at the end of this subsection, and that is the issue of normalization during the training process. In general, normalizing the input and output samples is an important step in gradient-based optimization techniques because it helps to ensure a more stable training process by maintaining balanced weight updates and trying to avoid exploding or vanishing gradients [55]. It also helps reduce overfitting and speed up convergence, especially when dealing with e.g. saturating nonlinearities in the hidden layers. There are several methods that could be used to normalize the data, e.g. minmax normalization, batch normalization, and layer normalization. In the model, normalizing the outputs means a simple scaling in the resulting path integrals; however, input normalization means that we will change the resulting frequencies and shifts in each of the quadratic kernels. This can be easily shown by taking e.g. the normalized output $y_N(T;\theta) = \mathcal{N}_y y(T;\theta)$ to the general expression for the response of the neural network in Eq. (31) (using the more compact notation shown in Eqs. (25)–(27)) to a normalized input $\mathcal{N}_u u(\tau)$ as:

$$y_N(T;\theta) = \mathcal{N}_y\Big[d_0 + \sum_{k=1}^{M} \hat{a}_k(T;\theta)\exp\Big(-\int_0^T d\tau\,\Big[A_k\Big(\mathcal{N}_u u(\tau) + \frac{B_k}{2A_k}\Big)^2\Big]\Big)\Big] =$$

$$= \mathcal{N}_y d_0 + \sum_{k=1}^{M} \mathcal{N}_y \hat{a}_k(T;\theta)\exp\Big(-\int_0^T d\tau\,\Big[A_k\mathcal{N}_u^2\Big(u(\tau) + \frac{B_k}{2A_k\mathcal{N}_u}\Big)^2\Big]\Big) \tag{39}$$

where $\mathcal{N}_y$ is a normalization to the output and $\mathcal{N}_u$ is the input normalization factor, while $A_k$, $B_k$, and $C_k$ are a combination of the free parameters of the neural network, as was introduced before. In the second line we took out the normalization factor from the square so that we arrive at the general quadratic form that can be used to redefine the frequencies and shifts of the harmonic oscillator kernels in Eq. (29) that approximates the original path integral as:

$$\omega_k^* = \sqrt{\frac{2A_k\mathcal{N}_u^2}{m}}, \tag{40}$$

$$s_k^* = -\frac{B_k}{2A_k\mathcal{N}_u}, \tag{41}$$

where the $\omega_k^*$ are the new frequencies, while $s_k^*$ are the new shifts that appeared because of the normalization of the input samples. Taking into consideration the effects of input–output normalization, the original path integral can be approximated as follows:

$$K(x_f, x_i, T) \approx \mathcal{N}_y d_0 F(x_f, x_i, T) + \sum_{k=1}^{M} \mathcal{N}_y \hat{a}_k(T) Q_k(x_f, x_i, T; \omega_k^*, s_k^*), \tag{42}$$

where $F(x_f, x_i, T)$ is the solution for the free-particle path integral, while $Q_k(x_f, x_i, T; \omega_k^*, s_k^*)$ is the solution for the shifted harmonic oscillator path integral with frequencies $\omega_k^*$ and shifts $s_k^*$.

Taking into consideration all the above-mentioned details, the optimization for the free parameters or, in other words, the training of the neural network can be done by following the next few steps:

- Determine the interval where the paths are giving dominant contributions by examining the response of the full system, including the kinetic terms.
- Generate training, validation, and test samples with the PCHIP (or any other) method in the predefined operating range.
- Normalize the inputs and the outputs.
- Train the network by a well suited method, e.g. gradient descent, by including the necessary constraints for the frequencies.

- Test the neural network on the generated test samples individually and by summing them as well, and determine the model errors.

Some remarks are in order regarding the last point, where we calculate the model error. Usually a neural network is trained by separating the data into training, validation, and test sets, where the optimization is done by using the training set and taking into consideration the response for the validation set so that there will be no overtraining and good generalization can be obtained. At the end, a test set is used to determine the model errors. In this case the individual errors are not meaningful because in the path integral we have to sum over a very large number of input paths; therefore, we need to estimate a global error including the response for many paths. If good generalization is achieved, this error will not "blow up", and we can estimate its statistical behavior for very large sets of data. The full training procedure with the model error estimation will be shown in detail through the examples in Section 4.

## 4. Results

In this section the neural network model will be applied to a purely real case and to a complex potential case as well, where the training procedure is also briefly described through the two examples. In the first example, the bound state energies and wave functions will be extracted through the large time behavior of the Euclidean propagator and compared to the numerical solutions, while in the second example, the real and imaginary parts of the propagators will be compared to the analytical solutions.

### 4.1. *Real potentials*

The physical problem that we would like to solve here is related to the quantum tunneling phenomena [56] in a double-well potential that can be described by the following Euclidean path integral:

$$K(x_f, x_i, T) = \int\limits_{\substack{x(T)=x_f \\ x(0)=x_i}} \mathcal{D}x(\tau) \exp\left(-\int_0^T d\tau \left(\frac{m}{2}\dot{x}^2 + V(x)\right)\right), \tag{43}$$

where $m$ is the mass of the particle, and the potential term for the double-well can be given as:

$$V(x) = \alpha x^4 - \beta x^2, \tag{44}$$

where we set the parameters to $\alpha = 0.05$ and $\beta = 1$. This potential describes a symmetric quantum mechanical system that has two local minima at $x = \sqrt{\frac{b}{2a}} = 3.16$ corresponding to its two stable equilibrium points, which are separated by a finite potential barrier. As the potential is purely real, the radial basis function expansion shown in Eq. (26) can be directly applied; therefore, the target outputs of the neural network model for an input $u_i(T)$ can be given in the following form:

$$y_{\text{true},i}(T) = \exp\left(-\int_0^T d\tau \left[\alpha u_i^4(\tau) - \beta u_i^2(\tau)\right]\right), \tag{45}$$

where $T$ is the Euclidean time, i.e. the upper limit until we have to integrate out the given polynomial function of the $u_i(\tau)$ training inputs. Extracting the bound state information from the Euclidean propagator means that we are primarily interested in the "large" Euclidean time behavior of the system; however, it is not straightforward to identify the behavior that we could consider "large" in this context. In practice, we could examine the Euclidean time dependence of the trace-normalized $K(x, x, T)$ propagator that should approximate the square of the normalized bound state wave function; therefore, if this quantity does not change anymore or the change is negligible, we could assume that the contributions of the excited states become negligible.

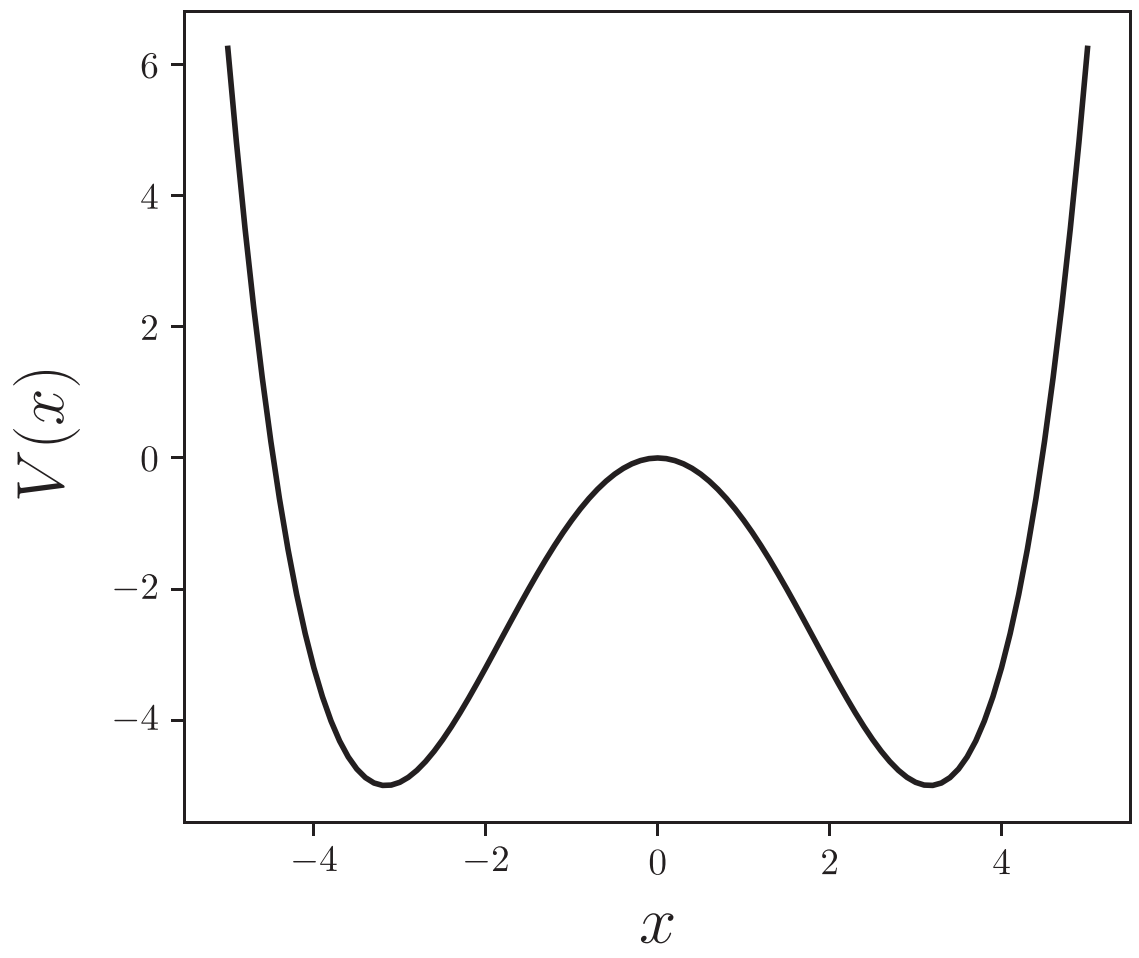

**Fig. 4.** The $V(x) = \alpha x^4 - \beta x^2$ double-well potential with parameters $\alpha = 0.05$ and $\beta = 1$, which will be used to test the neural network model for real valued potentials.

For our first test, let us consider the time interval of $T \in [0.5, 1]$ and set the mass to $m = 1$. To train the network, $N_{\text{train}} = 40\,000$ training and $N_{\text{valid}} = 10\,000$ validation samples were generated by the PCHIP method with the boundary conditions $x = x_i = x_f \in [-6, 6]$. This means that the operating range of the system, or in other words where we could make estimations for the propagator or the partition function, is in the range of $x \in [-6, 6]$. The range of values for the $u_i(\tau)$ input samples is chosen so that the generated paths in this region are dominant in the sense that if a path consists of many points outside this range, the contribution of the corresponding Boltzmann weight $e^{-S_E}$ will be negligible. This is related to the fact that the dominant paths will be the ones that are near to the classical paths, which depends on the $m$ mass and also on the parameters of the potentials; therefore, it has to be estimated in each case. In Fig. 4, the double-well potential that will be used to test the model is shown, where a preliminary guess for the interval where the dominant paths should lie could be given as $|u_i(\tau)| \sim [-6, 6]$, because outside of these limits the potential will give large positive values; thus, the probability that a particle stays in the outside region will be exponentially suppressed.

To see how the weight suppression acts through the input samples, a simple "minmax" analysis can be done as follows:

- Generate a large amount of samples $u_i(\tau)$, where $\tau$ goes from 0 to $T$.
- Calculate the full weight as:

$$w_i = \exp\left( -\int_0^T d\tau \left[ \frac{m}{2} \dot{u}_i^2(\tau) + \alpha u_i^4(\tau) - \beta u_i^2(\tau) \right] \right). \tag{46}$$

- Determine the maximum value of the weights as $M = \max(w_i)$.
- Select the samples that could give relatively large contributions, e.g. larger than 1% of the maximum value: $w_i \geq 0.01M$.
- Calculate the minimum and maximum values for the selected samples.

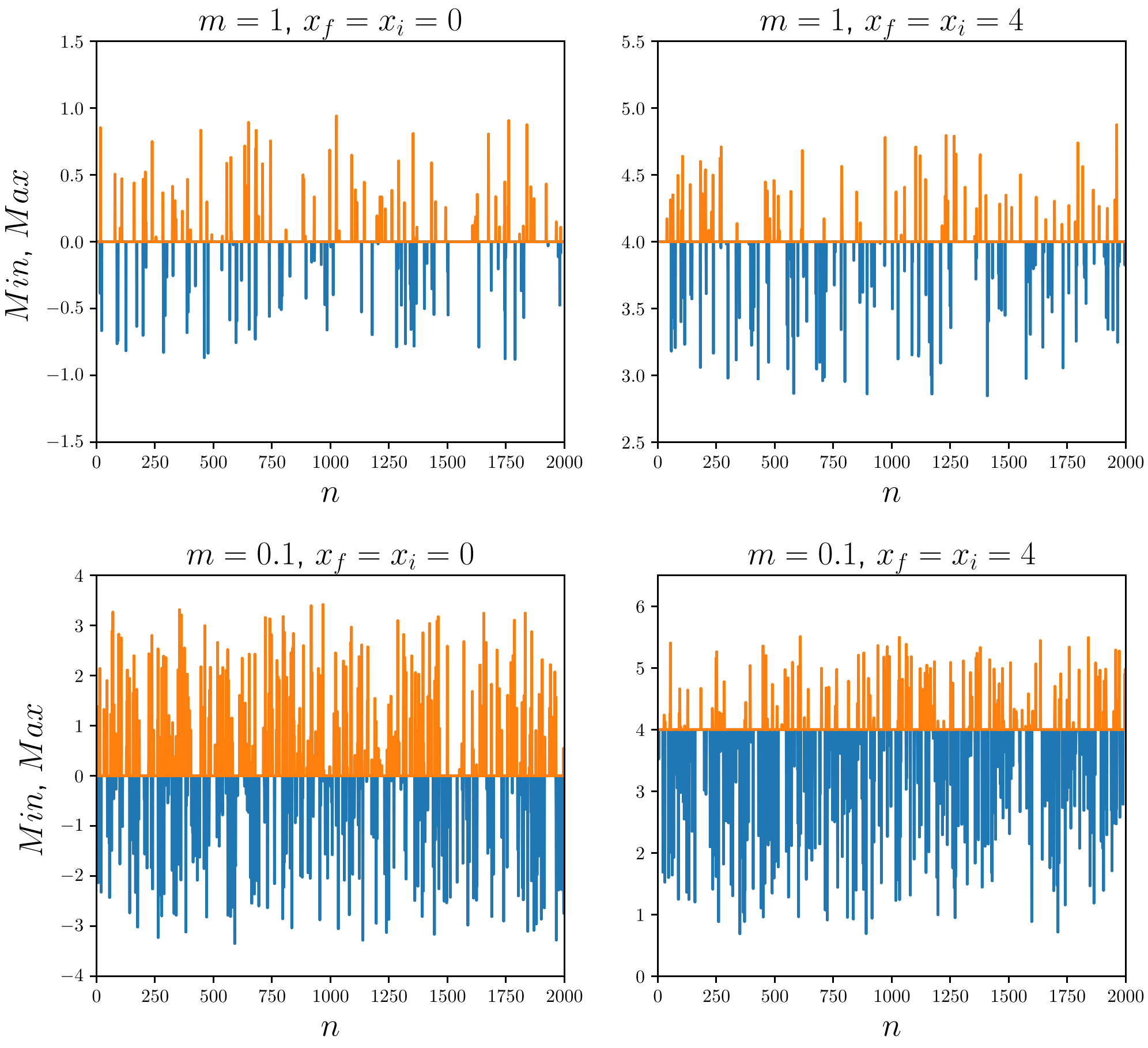

**Fig. 5.** The minimum (blue) and maximum (orange) values of the dominant paths for $m = 1$ and $m = 0.1$, with the boundary conditions $x_i = x_f = 0$ and $x_i = x_f = 4$, in the $\tau \in [0, 0.5]$ Euclidean time interval.

The minimum and maximum values of the selected paths will provide an estimation of the intervals where we should generate the samples. The naive expectation would be that this interval should be in the interval that could be guessed from the potentials alone; however, here we are also taking into consideration the kinetic part in the weight factor; thus it is a dynamical estimation of the possible paths. In Fig. 5, two cases are shown with the Euclidean time fixed to $T = 0.5$, where the left plot has been calculated by setting the boundaries to $x_f = x_i = 2$, while the right plot is calculated with the boundaries set to $x_f = x_i = 4$. The final samples were selected from 100 000 generated paths, which were generated by the PCHIP method in the interval $u_i(\tau) \in [-20, 20]$. The minimum and maximum values that are observed for the two cases with different masses suggest that the interval of the training samples should be larger for the lower mass case. The upper left plot shows that in the case of $m = 1$ with $x = 0$, we need to include paths that are between $x \pm 1$ to be able to consider the most dominant paths in the path integral. In the case of $m = 0.1$ and $x = 0$ (lower left plot), this interval becomes $x \pm 3$. The plots on the right side when $x = 4$ show an interesting but sensible asymmetric behavior, where the lower bounds show the same behavior as before, but now the upper bounds tend to be lower than 6, which is exactly the behavior that we expected just by looking at the shape of the potential. Because of the symmetric nature of the potential and therefore the corresponding weight factors, this behavior is similar in the case when we generate paths, e.g. with $x = -4$ as well, in which case the minimum of the dominant paths should be larger than $-6$. To conclude, if we want to describe the system with $m > 0.1$ masses at $T = 0.5$, then we have to generate paths that are between the interval $u(\tau) \in [\max(x-3, -6), \min(x+3, 6)]$ for all $\tau \in [0, 0.5]$. For other $T$ Euclidean times, or time intervals, the interval could be different, but this gives a general idea on how to estimate them by simply calculating the Boltzmann weights for each path.

The next step is the determination of the number of inputs and the corresponding centers of the inputs of the neural network, e.g. $z_i = \int d\tau\ (u_i(\tau) - c_i)^2$. To give a good cover of the input space, let us set up 7 inputs with the centers $c_i = [-5, -3, -1, 0, 1, 3, 5]$. It is not necessary to use symmetric centers; however, because of the symmetric nature of the potential, it could help in the training process. It is worth noting that the final centers of the $M$ quadratic kernels that we obtain through training the network will be different, and the $c_i$ values that we give here only serve as initial conditions that could help the network to converge faster.

The final ingredient that we should determine before we train the network is the number of neurons in the hidden layer, which will determine the number of quadratic kernels in the final expansion that is given by the output of the neural network. In general, there is no exact way to determine the number of necessary neurons in a neural network, and the usual way to do this is to train the network with different numbers of neurons and choose the one configuration that can describe the data well enough. Because of the fact that in the path integral we have to sum over a very large number of paths, the absolute error of an individual sample is not meaningful in itself when we want to characterize the total error of the model. On the other hand, the training process needs the absolute error (or any other loss function) to be able to optimize the model for the free parameters. Taking into consideration the mentioned problems, we separate the model validation from its training in the following way. The training goes as the usual backpropagation method using the absolute error as the loss function; however, to validate each model, we will use the following error definition:

$$E_A(N_{\text{test}}) = \frac{\frac{1}{N_{\text{test}}} \sum_{i=1}^{N_{\text{test}}} |y_{\text{est},i} - y_{\text{true},i}|}{\sqrt{\frac{1}{N_{\text{test}}} \sum_{i=1}^{N_{\text{test}}} (y_{\text{true},i} - \tilde{y}_{\text{true}})^2}}, \tag{47}$$

where $y_{\text{est},i}$ is the estimated output of the neural network for the $i$th test sample, $y_{\text{true},i}$ is the true output, and $N_{\text{test}}$ is the number of test samples that we use for the sum, while $\tilde{y}_{\text{true}}$ is the mean value of the target samples. By definition the denominator is the standard deviation of the true values; thus $E_A$ measures the averaged absolute errors that are normalized by the relative spread of the data. This definition makes it possible to measure the accumulated absolute error for a large finite sum without neglecting the possible cancellations due to the positive and negative pointwise differences. Additionally, the normalization by the standard deviation scales the error to a meaningful range that corresponds to the spread of the target data.

If the model error has good generalization and the error is approximately bias-free with some well defined statistical properties, e.g. finite variance, then $E_A(N_{\text{test}})$ should converge to a small finite value for large $N_{\text{test}}$. In the following tests, we have calculated the errors in the range of $N_{\text{test}} \in [1000, 15\,000]$ to see how $E_A$ behaves with increasing sample size. In Fig. 6 we show the training process in four different scenarios, with $M_1 = 1$, $M_2 = 5$, $M_3 = 10$, and $M_4 = 50$ neurons, and compare the results for the evolution of the training and validation losses. In each case the error $E_A$ for $N_{\text{test}} = 15\,000$ is also shown to be able to characterize the goodness of the model. From the comparison in Fig. 6, we can see that the $M = 1$ neuron case is not capable of giving a good estimation because of the large $E_A = 0.65$ error; however, in the case when the number of neurons in the hidden layer is greater than 10, the model is able to learn and generalize with much better accuracy. In addition to the aggregated error, it is also very important to check the actual pointwise differences and their distributions to make sure that the model is able to generalize uniformly in the operating range. In Fig. 7 the evolution of $E_A(N)$ for different sizes of test samples is shown for the case of $M = 50$ neurons, showing the convergence to a finite value, while on the left plot in Fig. 8, the true and estimated data are compared to-gether, and on the right plot, the normalized distribution of the differences is also shown. From these comparisons we see that the normalized aggregated error $E_A$ is capable of quantifying the goodness of the model. The histogram of the pointwise differences shows a very peaked distribution around zero, which is consistent with the previous statement about the convergence of $E_A(N_{\text{test}})$ for large $N_{\text{test}}$. Additionally, we also have to consider the fact that in these cases we only trained the network for one specific Euclidean time $T = 0.5$. In the case when we want to train the network to an interval of Euclidean times, the problem becomes a bit harder, and it has to be analyzed separately. The estimations using only one Euclidean time can be used as a first estimation on what to expect, but for the interval $T \in [0.5, 1]$ we have to reanalyze the errors. Taking into consideration all the above, we will train two neural networks with different numbers of hidden layers ($M = 10$ and $M = 50$), each having the following parameters:

- Number of inputs: $N = 7$, with centers $c_i = [-5, -3, -1, 0, 1, 3, 5]$.
- Number of neurons in the hidden layer: $M = 10$ and $M = 50$.
- Euclidean time interval: $T \in [0.5, 1]$, generated uniformly for each training sample.
- Interval of the boundary values: $x \in [-6, 6]$, where $x = x_i = x_f$, generated randomly from a uniform distribution in each sample.
- Interval of the training samples (paths): $u(\tau) \in [\max(x - 3, -6), \min(x + 3, 6)]$, for every $\tau \in [0, T]$.

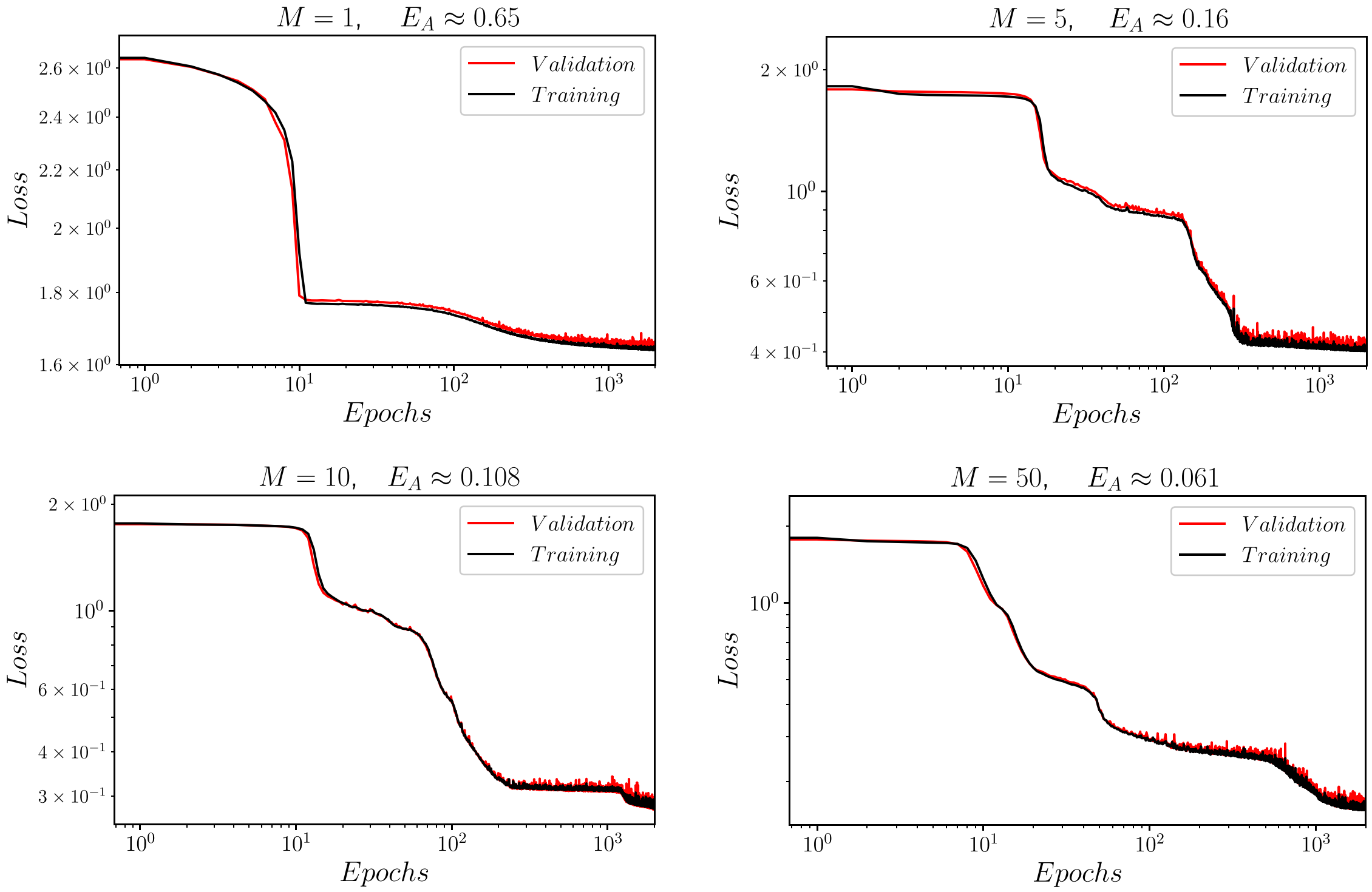

**Fig. 6.** Evolution of the training and validation losses of 4 different neural networks with $M_1 = 1, M_2 = 5, M_3 = 10$, and $M_4 = 50$ neurons in the hidden layers for a system where the Euclidean time is set to $T = 0.5$. The errors $E_A$ are estimated by using 15 000 test samples.

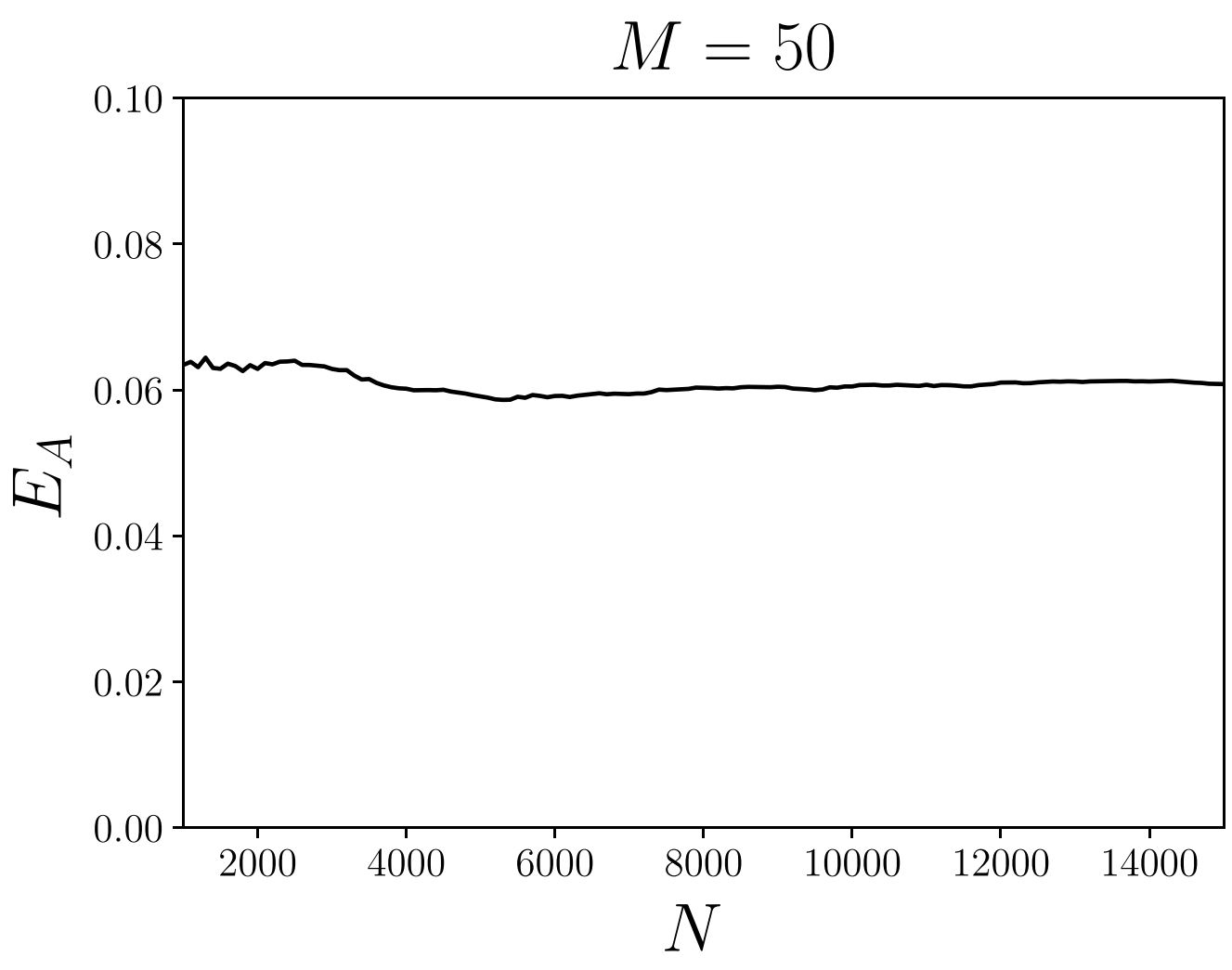

**Fig. 7.** The $E_A$ error dependence on the number of test samples that is calculated by Eq. (47) using the neural network with $M = 50$ neurons in the hidden layer for the system with $T = 0.5$.

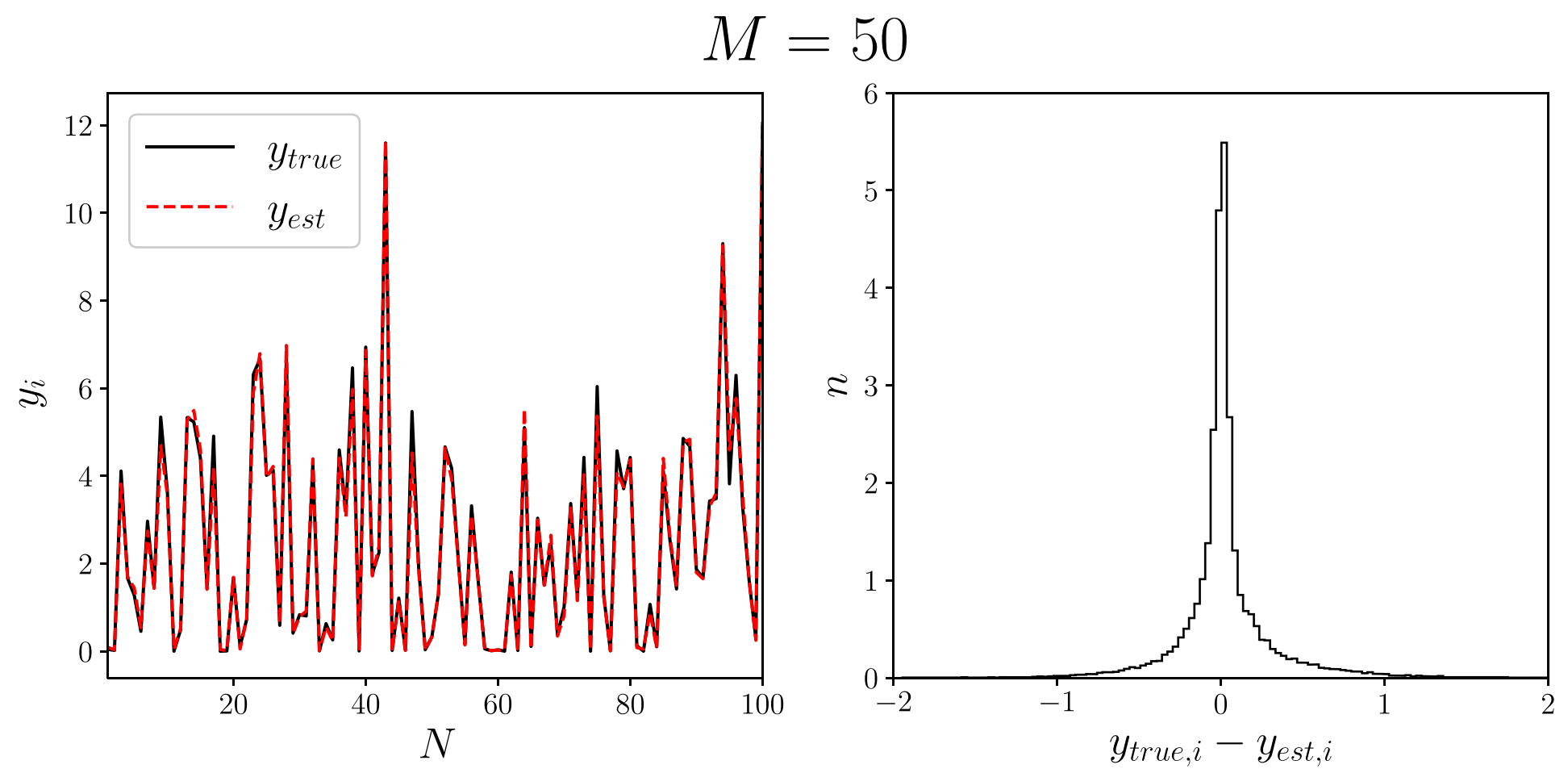

**Fig. 8.** Comparison of the true values and the approximations for 100 pieces of test data (left), and the normalized distribution of their differences (right) using $N_{\text{test}} = 15\,000$ test data for the $M = 50$ neuron model.

The evolution of the unnormalized absolute loss for the training and validation sets for the two networks is shown in Fig. 9, while the $E_A$ error dependence on the number of test samples can be seen in Fig. 10.

From the obtained $E_A$ values and from the evolution of the training and validation errors, it can be seen that a better approximation could be achieved by having more neurons and/or letting the training run further (more epochs), but for our purpose of seeing a rough evolution of the propagator, the network with $M = 50$ neurons in the hidden layer that has $E_A = 0.1$ will suffice. To see how the propagator behaves in the given Euclidean time interval, we have calculated the time evolution of the trace-normalized propagator $K_N(x, T)$ with the second network, where $M = 50$, with a slightly lower error than in the $M = 10$ case, in the range of $x \in [-6, 6]$ and $T \in [0.5, 1]$. The results can be seen in Fig. 11.

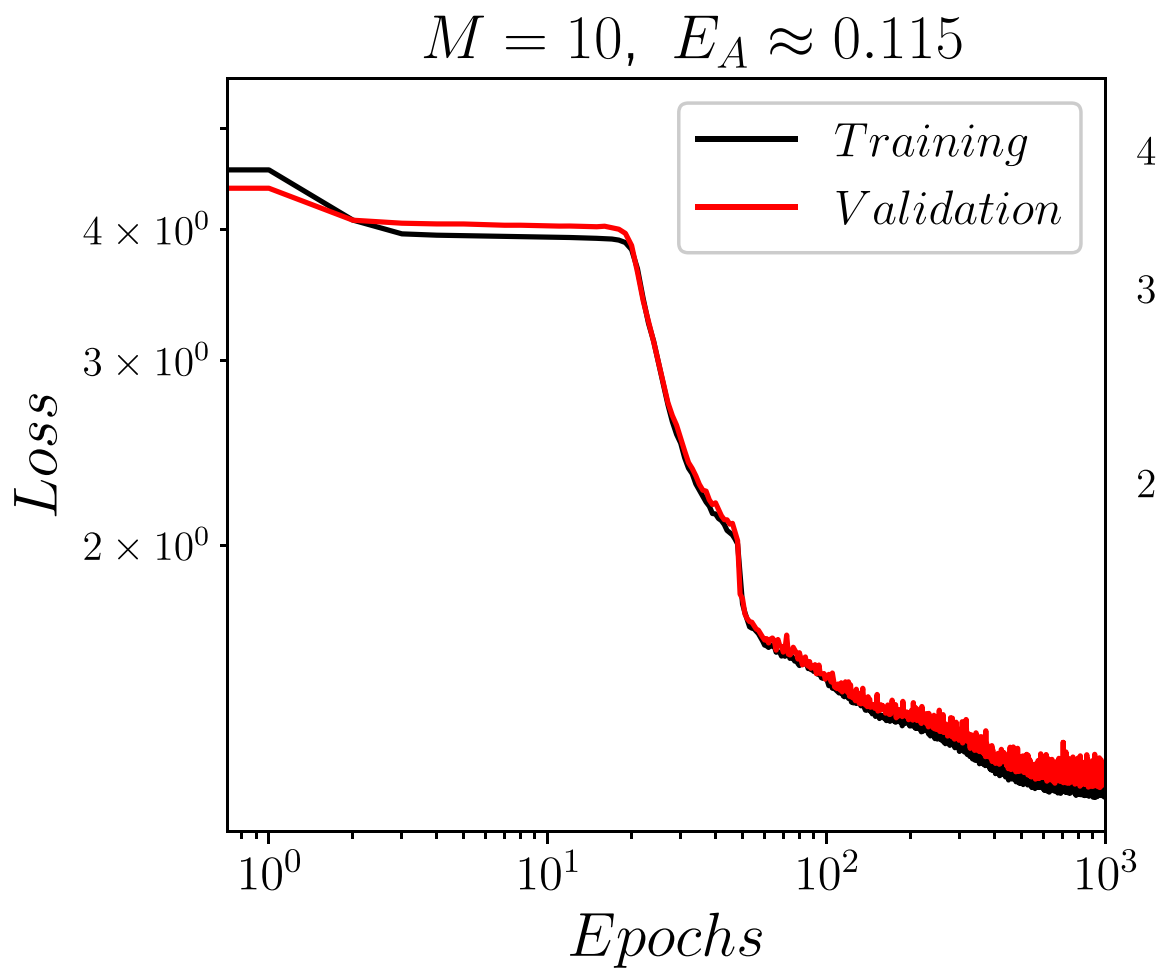

**Fig. 9.** Evolution of the training and validation losses of two neural networks with $M = 10$ and with $M = 50$ neurons in the hidden layer, for the case of a $T \in [0.5, 1]$ Euclidean time interval.

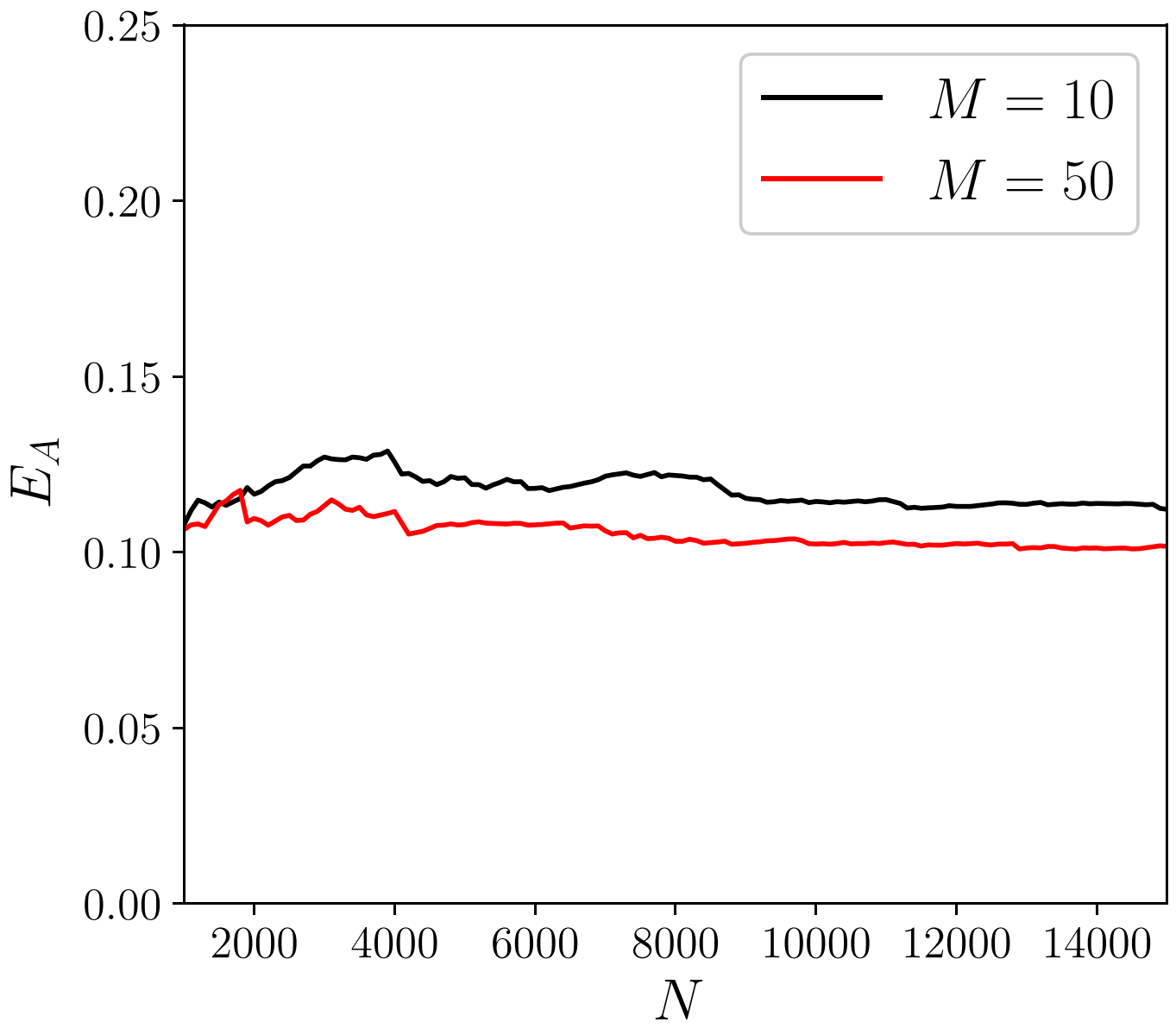

**Fig. 10.** The evolution of $E_A$ for different number of test samples in the case of $M = 10$ and $M = 50$ neurons, and a $T \in [0.5, 1]$ Euclidean time interval.

From the propagator it can be deduced that $T = 1$ is not large enough to extract bound state information because the trace-normalized propagator is still changing as we go forward in Euclidean time; therefore, we need to extend or change the operating range of $T$. To do this, let us change the time interval to $T \in [2.5, 3]$, keeping the other parameters (number of inputs, centers, number of neurons $M = 50$, boundaries, and the interval of the training samples) the same, and train the neural network in the same way as before. The results of the training and the model validation can be seen in Fig. 12, where, according to the errors $E_A \approx 0.018$, a very good generalization is achieved.

The time evolution of the trace-normalized Euclidean propagator can be followed in Fig. 13,

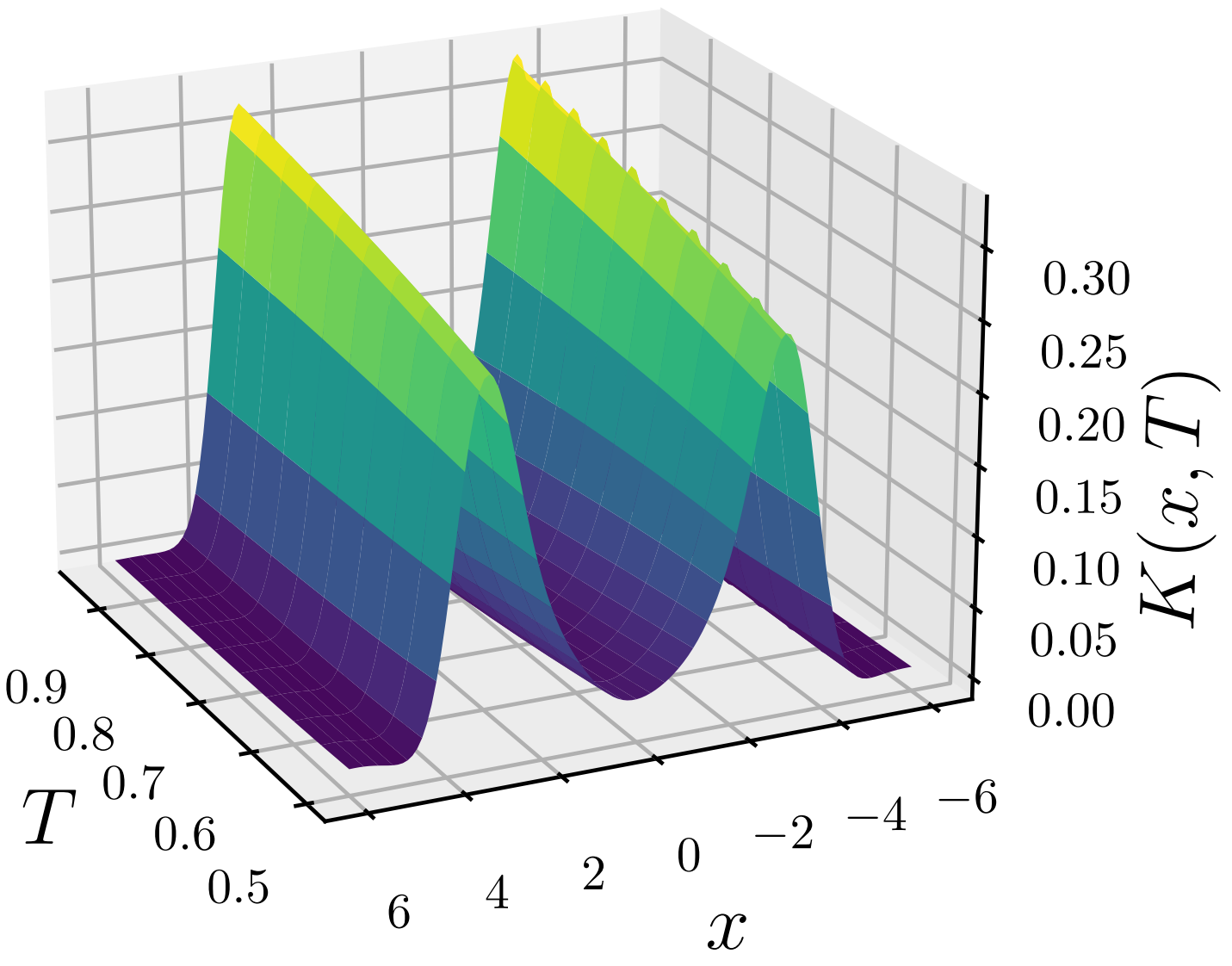

**Fig. 11.** Euclidean time ($T$) and position ($x$) dependence of the trace-normalized propagator $K(x, T)$, extracted from the trained neural network that has $M = 50$ neurons in the hidden layer, shown in the time interval $T \in [0.5, 1]$. As the propagator is still changing in time, this interval is not enough to extract the bound state wave function.

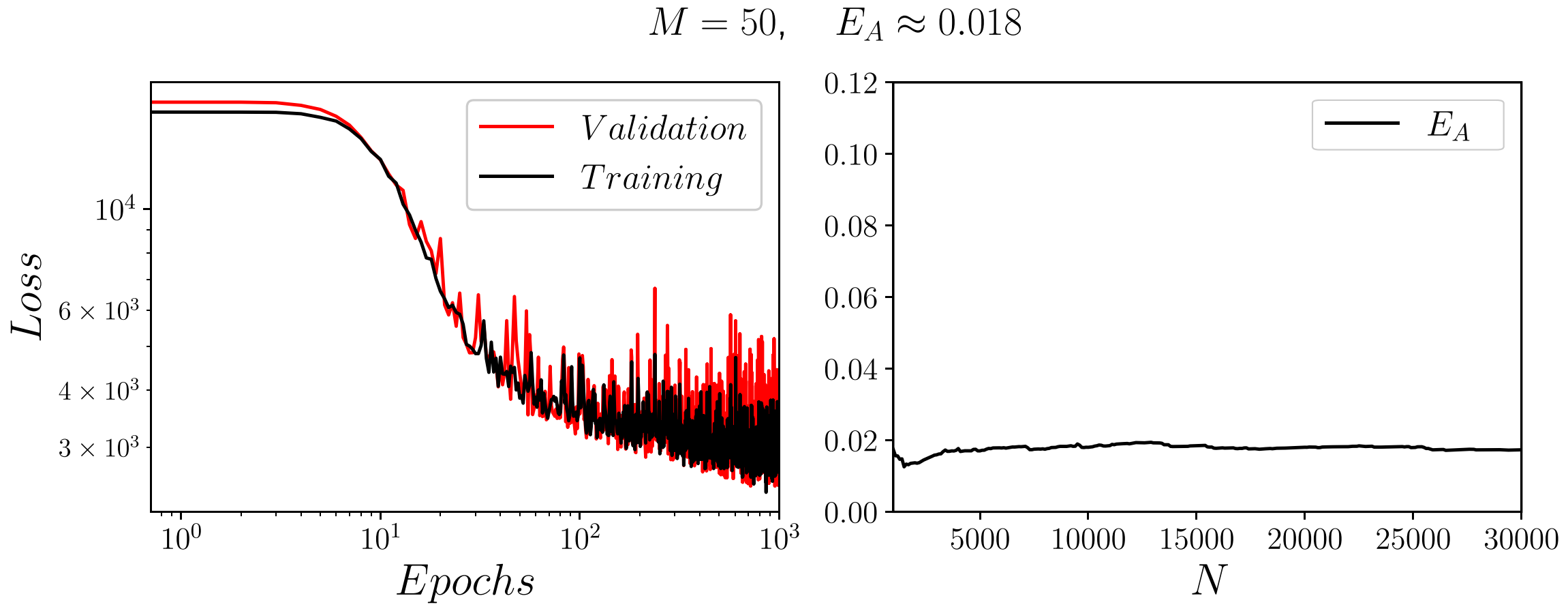

**Fig. 12.** Evolution of the training and validation losses for the system with $M = 50$ neurons trained in the Euclidean time interval $T \in [2.5, 3]$. On the right side, the $E_A$ error dependence on the number of test samples is shown for the same network.

which shows an almost time-independent behavior; therefore, this means that we can extract the bound state energy by fitting a linear function to the logarithm of the partition function.

In Fig. 14 the true bound state wave function, obtained by solving the Schrödinger equation, is compared to the normalized propagator that is calculated from the trained neural network model through Eq. (29) after extracting the trained parameters from the neural network model in Eq. (29).

The model error is estimated through an ensemble-based method, where we have generated a large set of neural networks, each with a different number of hidden layers between $M = 40$ and $M = 50$, and trained them with different sets of training and validation samples. Using the obtained propagators at each $x$ and $T$, we determined the distribution of $K(x, T)$ at each $(x, T)$ coordinate, from which the approximation could be given as the mean of these values ± their standard deviation. The actual values have been calculated using 100 different configurations and training/validation samples.

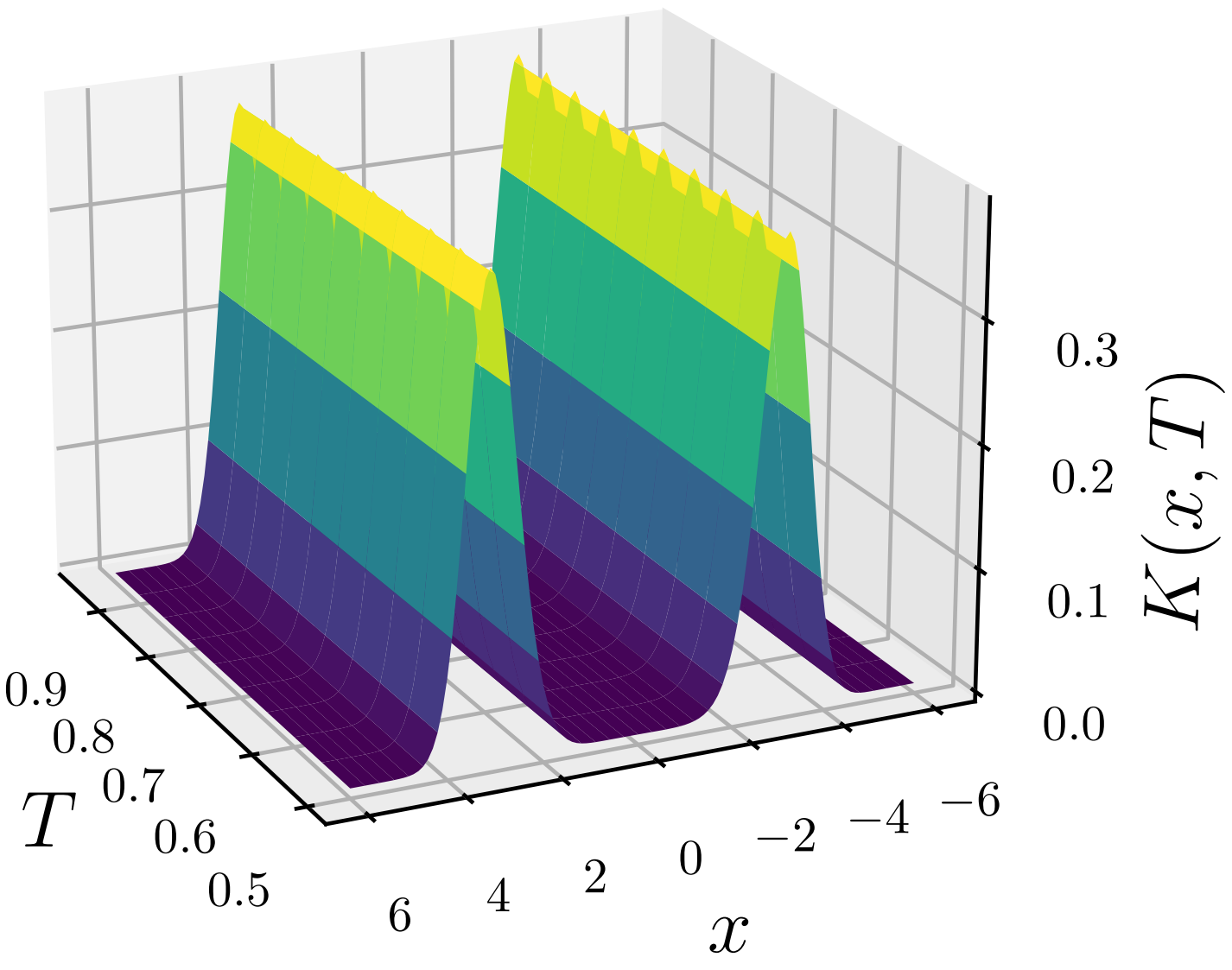

**Fig. 13.** Euclidean time ($T$) and position ($x$) dependence of the trace-normalized propagator $K(x, T)$, extracted from the trained neural network with $M = 50$ neurons, shown in the time interval $T \in [2.5, 3]$. As the normalized propagator is approximately time independent, it is possible to extract the bound state wave function.

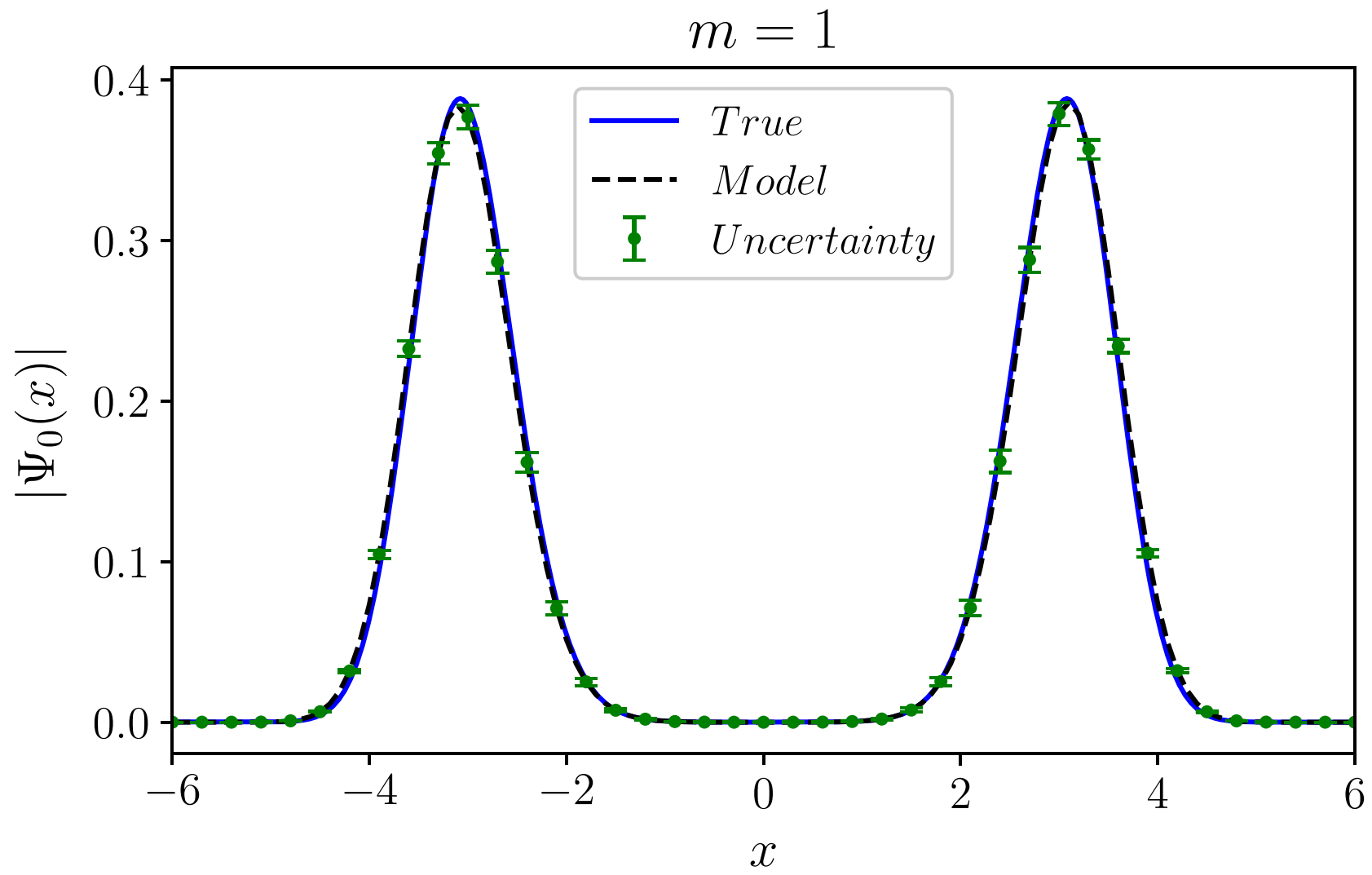

**Fig. 14.** Bound state wave function for a particle with $m = 1$ in a double-well potential, compared to the trace-normalized propagator at large Euclidean time $T = 3$ that is extracted from the neural network model.

The estimated bound state wave function that is obtained by the large Euclidean time propagator is very close to the real one that is obtained by solving the Schrödinger equation, which is the expected result considering the estimated uncertainty of the model. One of the great advantages of this model is the ability to give a continuous, analytically calculable result for the propagator in the operating range that is defined through the training of the network.

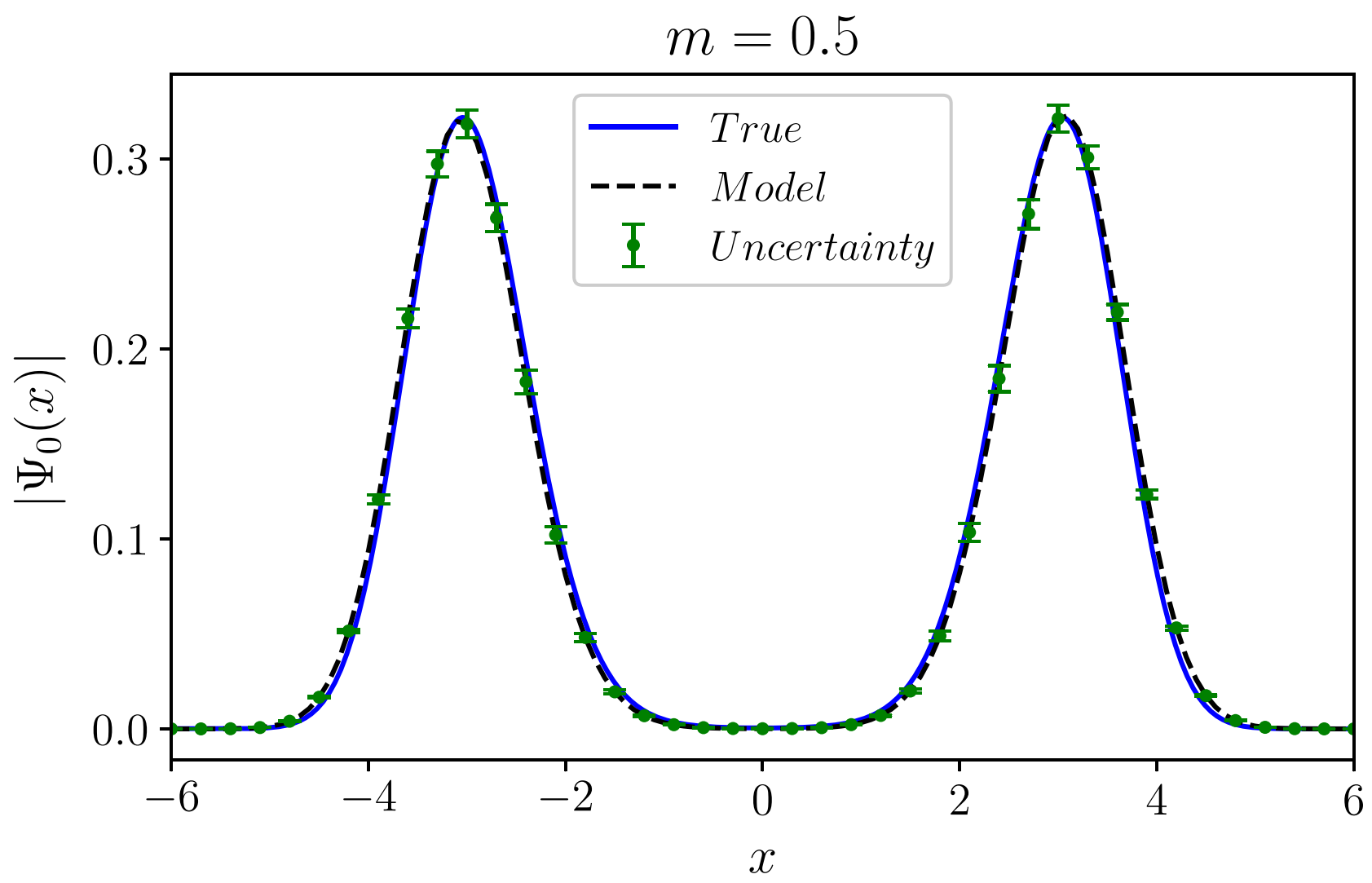

**Fig. 15.** Bound state wave function for a particle with $m = 0.5$ in a double-well potential, compared to the trace-normalized propagator at large Euclidean time $T = 3$ that is extracted from the neural network model.

In this case we have a closed-form expression for the propagator $K(x, T)$ in the range of $x, T \in [-6, 6]_x \times [2.5, 3]_T$ (and for $x, T \in [-6, 6]_x \times [0.5, 1]_T$ in the previous example), which can be used to integrate out in $x$ to obtain an analytical form of the $Z(T)$ partition function in the given Euclidean time interval. One has to be careful, however, to consider the limits of the integral according to the operating range of the model; therefore, instead of $\int_{-\infty}^{\infty} dx$, the integral $\int_{x_{\min}}^{x_{\max}} dx$ has to be calculated. This also means that we have to make sure that the propagator at the boundaries is close to zero so that the integral is meaningful. If this condition is not satisfied, then one can extend the boundaries of the model and train the network again or, in some cases, extrapolate down to zero with some given functional form. In this case the propagator dies out well before it reaches the boundaries, so we do not have to consider this problem.

To finish this subsection, let us show that the model is also capable of giving the correct propagator to other masses as well without any modification or retraining of the model. The neural network does not contain any information about the masses directly; however, through the frequency terms in the harmonic oscillator kernels, e.g. in Eq. (36), and through the interval of the training samples, the information on the masses is encoded indirectly, thus giving an operating range for the masses as well. This usually means a lower limit on the masses, where the kinetic term becomes less important and the potential term dominates more and more. This, of course, depends on the specific type of Lagrangian that we consider, and it is a modeling step on how to include the masses in the neural network. In this case the interval of the training samples was chosen so that the model should be capable of including masses down to $m \gtrsim 0.1$.

In Fig. 15, the estimated bound state wave function in the case of $m = 0.5$ is calculated and compared to the numerical solution of the Schrödinger equation. In the case of the lower mass, the two peaks are somewhat broadened, and the corresponding maximum values at the peaks are lower than in the case of $m = 1$, behavior that is perfectly captured by the neural network model.

In the next section we will show how the neural network model could be used to describe systems with complex potentials as well, which could be important in describing, e.g. dissipative systems in quantum mechanics or systems at finite densities in quantum field theories.

### 4.2. *Potentials with imaginary parts*

Many interesting real-world phenomena that can be described by relativistic or nonrelativistic quantum physics could include dissipative effects, non-Hermitian dynamics, and finite density effects that essentially need the inclusion of complex potentials or interaction terms in the Lagrangians that describe the corresponding systems [57]. In quantum scattering, complex potentials could be used to describe resonant states and particle decay, while in open quantum systems [58], where the system interacts with its surroundings, imaginary parts in the optical potential could lead to dissipative effects, e.g. where the energy or total probability is not conserved. In the Euclidean path integral formulation of quantum chromodynamics at finite densities, the inclusion of the chemical potential associated with a corresponding conserved charge naturally leads to an imaginary part in the Lagrangian that leads to the famous "sign problem" in numerical lattice field theory.

In this paper, we only consider nonrelativistic quantum mechanics where the potential term will have an additional imaginary part as:

$$V(x) = V_R(x) + iV_I(x), \tag{48}$$

where $V_R(x)$ is the real part, $V_I(x)$ is the imaginary part of the potentials, while $i$ represents the imaginary unit. Introducing this potential into the path integral and using the fact that $e^{iV} = \cos(V) + i\sin(V)$, we can arrive at the following form:

$$\begin{aligned} K(x_f, x_i, T) &= \int\limits_{\substack{x(T)=x_f \\ x(0)=x_i}} \mathcal{D}x(\tau) \exp\left(-\int_0^T d\tau \left(\frac{m}{2}\dot{x}^2 + V_R(x) + iV_I(x)\right)\right) \\ &= \int\limits_{\substack{x(T)=x_f \\ x(0)=x_i}} \mathcal{D}x(\tau)\ e^{-\int_0^T d\tau \left(\frac{m}{2}\dot{x}^2 + V_R(x)\right)} \left[\cos\left(\int_0^T d\tau V_I(x)\right) - i\sin\left(\int_0^T d\tau V_I(x)\right)\right], \end{aligned} \tag{49}$$

where we have used the $\cos(-x) = \cos(x)$ and $\sin(-x) = -\sin(x)$ properties of the harmonic functions to obtain the final expression. Looking at the path integral, it is apparent that the propagator is separated into purely real and purely imaginary parts as:

$$K(x_f, x_i, T) = K_R(x_f, x_i, T) - iK_I(x_f, x_i, T), \tag{50}$$

where the real part can be given by:

$$K_R(x_f, x_i, T) = \int\limits_{\substack{x(T)=x_f \\ x(0)=x_i}} \mathcal{D}x(\tau) \left[\cos\left(\int_0^T d\tau V_I(x)\right)\right] e^{-\int_0^T d\tau \left(\frac{m}{2}\dot{x}^2 + V_R(x)\right)}, \tag{51}$$

and the imaginary part of the propagator can be expressed as:

$$K_I(x_f, x_i, T) = \int\limits_{\substack{x(T)=x_f \\ x(0)=x_i}} \mathcal{D}x(\tau) \left[ \sin\Big( \int\limits_0^T d\tau V_I(x) \Big) \right] e^{-\int\limits_0^T d\tau \left( \frac{m}{2}\dot{x}^2 + V_R(x) \right)}, \quad (52)$$

where the $V_R(x)$ real and $V_I(x)$ imaginary parts of the potential both appear in the real and in the imaginary parts of the propagator, where the imaginary part introduces an oscillating term, while the real part corresponds to the usual Boltzmann weight factor. The full path integral in this separated form introduces two real-valued path integrals that have to be solved, and while the oscillating terms could make the usual numerical calculations harder, the neural network model is well suited to solve this problem as it does not depend on a positive definite, well behaved weight factor to be able to generate the paths that have to be summed.

To extend the model for complex potentials, let us start from the already separated real and imaginary parts of the propagator in Eqs. (51) and (52). According to Eq. (33), we can write down the following two targets that have to be approximated by the neural network model:

$$y^{(1)}_{\text{true}}(T) = \sin\Big( \int\limits_0^T d\tau V_I(x) \Big) \exp\Big( - \int\limits_0^T d\tau V_R(x) \Big), \quad (53)$$

$$y^{(2)}_{\text{true}}(T) = \cos\Big( \int\limits_0^T d\tau V_I(x) \Big) \exp\Big( - \int\limits_0^T d\tau V_R(x) \Big), \quad (54)$$

where, just as before, we have separated the kinetic terms from the interactions, which in this case contain the contributions from the imaginary part of the potential as well. As we arrive at two distinct outputs that have to be approximated, it would be possible to consider them as separate path integrals and train two separate networks, where one describes the real part, while the other one describes the imaginary part of the full path integral. In the case when we have two separate neural networks according to the architecture shown in Fig. 2, we will have twice as many free parameters as in the purely real potential case. While it is perfectly possible to train two neural networks for the real and imaginary parts, it would be much more efficient if we could describe the full path integral with only one neural network, therefore giving a more robust approximation to the original problem.

To be able to estimate the real and imaginary parts of the propagator simultaneously, let us introduce a multiple input–multiple output (MIMO) system, which has a topology that is shown in Fig. 16.

The corresponding outputs of this neural network can be given in the following closed form:

$$y_j(T) = d_{0j} + \sum_{k=1}^{M} a_{kj} \exp\Big( - d_k - \sum_{i=1}^{N} \Big[ b_{ik} \int\limits_0^T d\tau \, (x - c_i)^2 \Big] \Big), \quad (55)$$

with $j = (1, 2)$, where $j = 1$ represents the real part of the path integral, while $j = 2$ represents the imaginary part. In this case the only change to the original formulation is the extension of the weight and bias parameters in the final (linear) layer by $a_k \to (a_{k1}, a_{k2})$ and $d_0 \to (d_{01}, d_{02})$. This change ultimately means that we have introduced complex linear weight factors and biases into the quadratic expansion of the path integral shown in Eq. (28), where both the $d_0$ bias term and the $\hat{a}_k(T)$ coefficients are complexified as:

$$d_0 = d_{01} - i d_{02}, \quad (56)$$

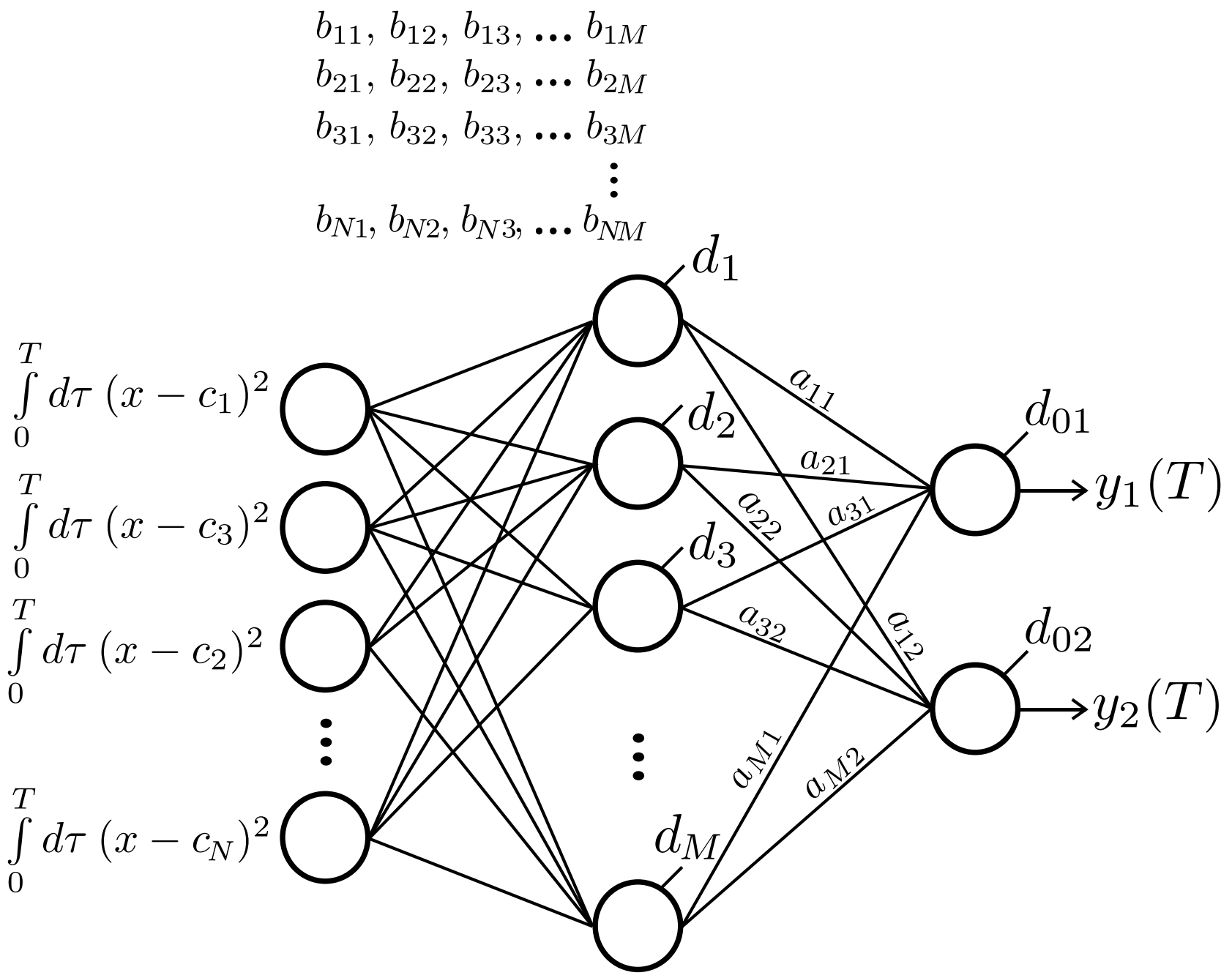

**Fig. 16.** A schematic view of the neural network model that is used to approximate path integrals with complex potentials, where $y_1(T)$ represents the real part, while $y_2(T)$ represents the imaginary part of the interaction terms.

$$\hat{a}_k(T) = (a_{k1} - ia_{k2})\exp\left(-d_k - \left[C_k - \frac{B_k^2}{4A_k}\right]T\right), \tag{57}$$

where the $A_k = \sum_{i=1}^N b_{ik}$, $B_k = -2\sum_{i=1}^N b_{ik}c_i$, and $C_k = \sum_{i=1}^N b_{ik}c_i^2$ parameters stay the same as they were before. Putting the coefficients back into Eq. (29), the full propagator can be estimated by the trained network parameters as follows:

$$\begin{aligned} K(x_f, x_i, T) \approx & \left[d_{01}F(x_f, x_i, T) + \sum_{k=1}^M a_{k1}\exp\left(-d_k - \left[C_k - \frac{B_k^2}{4A_k}\right]T\right)Q_k(x_f, x_i, T)\right] \\ & -i\left[d_{02}F(x_f, x_i, T) + \sum_{k=1}^M a_{k2}\exp\left(-d_k - \left[C_k - \frac{B_k^2}{4A_k}\right]T\right)Q_k(x_f, x_i, T)\right], \end{aligned} \tag{58}$$

where $Q_k(x_f, x_i, T)$ is the propagator of the shifted harmonic oscillator with mass $m$, frequency $\omega_k = \sqrt{2A_k/m}$, and shift $s_k = -\frac{B_k}{2A_k}$, while $F(x_f, x_i, T)$ is the free-particle propagator. Comparing Eq. (50) with Eq. (58), the first term in the large [.] brackets is the $K_R(x_f, x_i, T)$ real part, while the term in the second large [.] brackets is the $K_I(x_f, x_i, T)$ imaginary part of the full propagator.

By estimating the path integral in this way, the real and imaginary parts of the approximation will be more consistent with each other, and the model will represent the system as a whole and not as separate path integrals. The number of trainable parameters in this case will be:

$$N_\theta = 2 + N + 3M + M \times N, \tag{59}$$

where we have an additional bias in the output layer and $M$ additional linear weights that connect the outputs of the hidden layer to the second neuron in the output layer.

It is worth mentioning that in the case where the oscillating parts in the path integral are more dominant, e.g. the potentials have very large imaginary parts with less damping, it would be better if the neural network also contained oscillating terms, which could be achieved easily by complexifying the frequencies through the $b_{ik}$ parameters by letting $b_{ik} = b_{ik}^{(1)} + ib_{ik}^{(2)}$. The square of the new complex frequencies in this case could be given by:

$$\omega_k^2 = 2\frac{\sum_{i=1}^N b_{ik}^{(1)}}{m} + 2i\frac{\sum_{i=1}^N b_{ik}^{(2)}}{m}, \tag{60}$$

where $b_{ik}^{(1)}$ are the real and $b_{ik}^{(2)}$ the complex parts of the $b_{ik}$ weight parameters that connect the inputs to the neurons in the hidden layer. There are other possibilities for how to model such problems even without directly introducing new parameters; e.g. if we do not include the positivity constraint $\sum_i b_{ik} > \epsilon$ in Eq. (37), then the corresponding frequencies $\omega_k = \sqrt{\frac{2\sum_i b_{ik}}{m}}$ could be imaginary as well, giving oscillating terms in the approximation of the path integral. It is again a modeling step on how to choose the best method that is suitable for the actual physical problem at hand.

To test the method, let us use an analytically tractable quadratic problem, namely the harmonic oscillator with complex frequencies, and define the Euclidean path integral as follows:

$$K(x_f, x_i, T) = \int\limits_{\substack{x(T)=x_f \\ x(0)=x_i}} \mathcal{D}x(\tau) \exp\Big(-\int_0^T d\tau \left[\frac{m}{2}\dot{x}^2 + \frac{m}{2}\omega_R^2 x^2 + i\frac{m}{2}\omega_I^2 x^2\right]\Big), \tag{61}$$

where the complex frequency of the system is defined as $\omega^2 = \omega_R^2 + i\omega_I^2$. By separating the interaction from the kinetic terms and then using Euler's rule, the target outputs for a $u_i(\tau)$ path can be expressed as follows:

$$y_{\text{true},i}^{(1)}(T) = \sin\Big(\int_0^T d\tau \frac{m}{2}\omega_I^2 u_i^2(\tau)\Big) \exp\Big(-\int_0^T d\tau \frac{m}{2}\omega_R^2 u_i^2(\tau)\Big), \tag{62}$$

$$y_{\text{true},i}^{(2)}(T) = \cos\Big(\int_0^T d\tau \frac{m}{2}\omega_I^2 u_i^2(\tau)\Big) \exp\Big(-\int_0^T d\tau \frac{m}{2}\omega_R^2 u_i^2(\tau)\Big), \tag{63}$$

where we have kept the $m$ mass as a parameter in the interaction terms. Previously, in the case of the double-well potential, the mass parameter only affected the kinetic terms directly; therefore, we did not have to consider it during training. However, in this case, the mass directly affects the potential terms as well; thus, it has to be taken into consideration in the generation of the training samples. Because of the flexibility of the neural network, it is possible to extend the model so that it is capable of including external parameter dependence as well, e.g. in the case of the complex harmonic oscillator $K(x, T; m, \omega_R, \omega_I)$, or in the double-well case $K(x, T; m, a, b)$, where $a$ and $b$ are the coefficients of the double-well potential, in which case we need to introduce new path-independent inputs to the network. Note that in the case of the double-well potential, the mass dependence was indirectly taken into consideration as the interaction term did not depend on it. However, the operating range set by the interval of the generated paths depends on the mass of the system throughout the kinetic terms in the exponential weight factor.

By introducing new parameter-dependent inputs, the output of the network changes, but the general quadratic structure on the paths does not; therefore, it can be used as before to approx-imate the path integral through the analytical solution of shifted harmonic oscillators and the free propagator. As the main goal in this section is the description of path integrals with complex potentials and not the external parameter dependence, we will not consider the extension of the network with additional parameter dependence and will set the mass parameter as a fixed value throughout the training process.

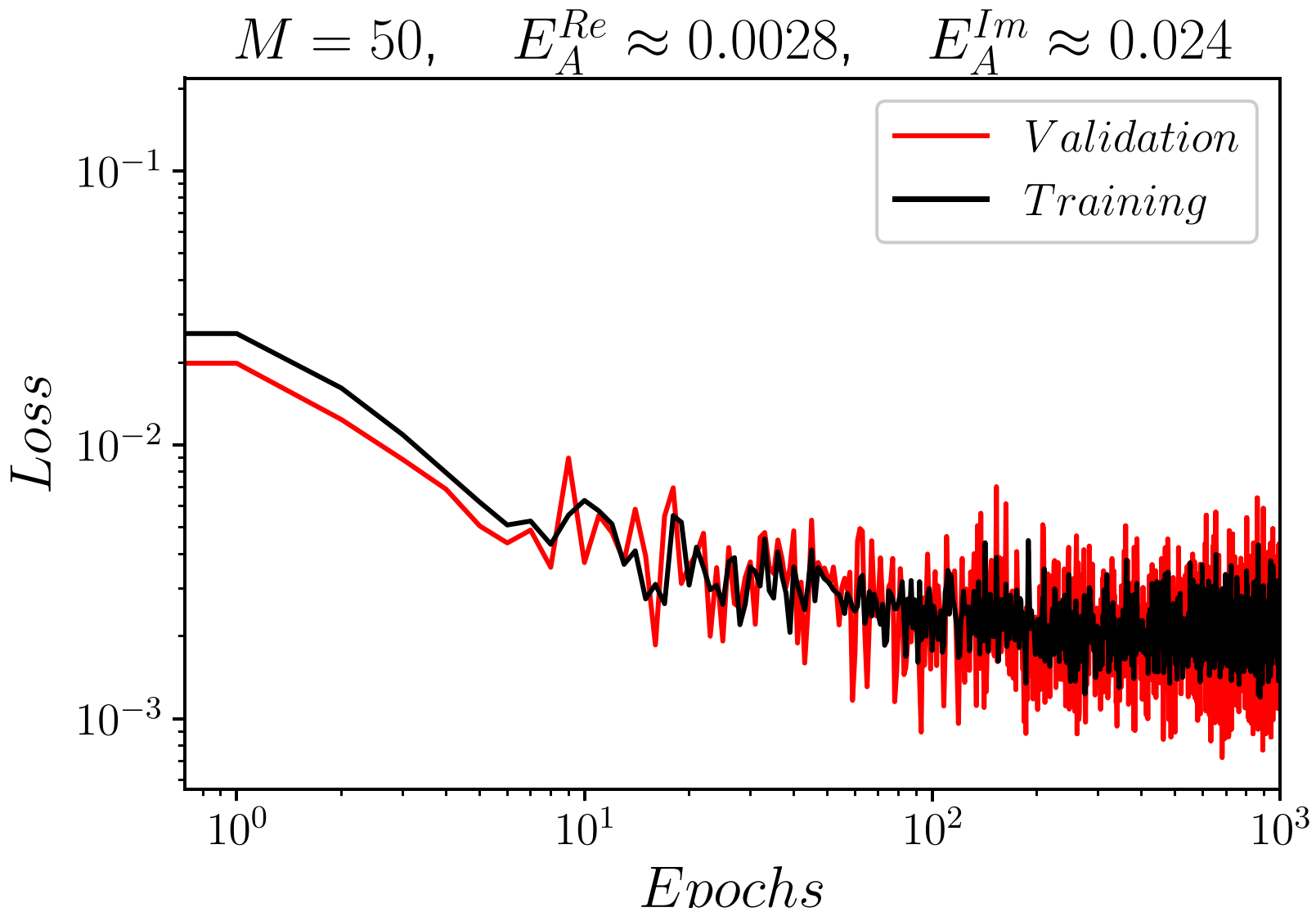

**Fig. 17.** Evolution of the training and validation losses during the training process for the harmonic oscillator path integral with complex frequencies, trained in the $T \in [0.5, 1]$ Euclidean time interval with a neural network that contains $M = 50$ neurons in the hidden layer.

To test the method, let us set up the model and the neural network that is shown in Fig. 16 as follows:

- Set mass to $m = 1$.
- Set the frequencies to $\omega_R = 0.5$ and $\omega_I = 0.25$.
- Number of inputs: $N = 7$, with centers $c_i = [-5, -3, -1, 0, 1, 3, 5]$.
- Number of neurons in the hidden layer: $M = 50$.
- Euclidean time interval: $T \in [0.5, 1]$.
- Interval of the boundary values: $x \in [-5, 5]$.
- Interval of the training samples (paths): $u(\tau) \in [\max(x-3, -6), \min(x+3, 6)]$, for every $\tau \in [0, T]$.

The neural network topology that we will use is the one shown in Fig. 16, where the real and imaginary parts are estimated by the two outputs $y_1(T)$ and $y_2(T)$, which means that the $a_k$ and $d_0$ parameters are complexified as $a_k \to a_{k1} - ia_{k2}$, and $d_0 \to d_{01} - id_{02}$.

To train the network, 50 000 training samples have been generated by the PCHIP method, 10 000 samples were generated for validation purposes, and the calculation of the $E_A^{\mathrm{Re}}$ and $E_A^{\mathrm{Im}}$ errors is done up to 15 000 test samples. The evolution of the training and validation losses is shown in Fig. 17.

The very small values of the $E_A^{\mathrm{Re}}$ and $E_A^{\mathrm{Im}}$ errors suggest that the overall uncertainty of the model will also be very small, which indeed can be seen in Fig. 18, where a comparison of the real and imaginary parts of the approximated and true propagators can be followed at $T = 1$.

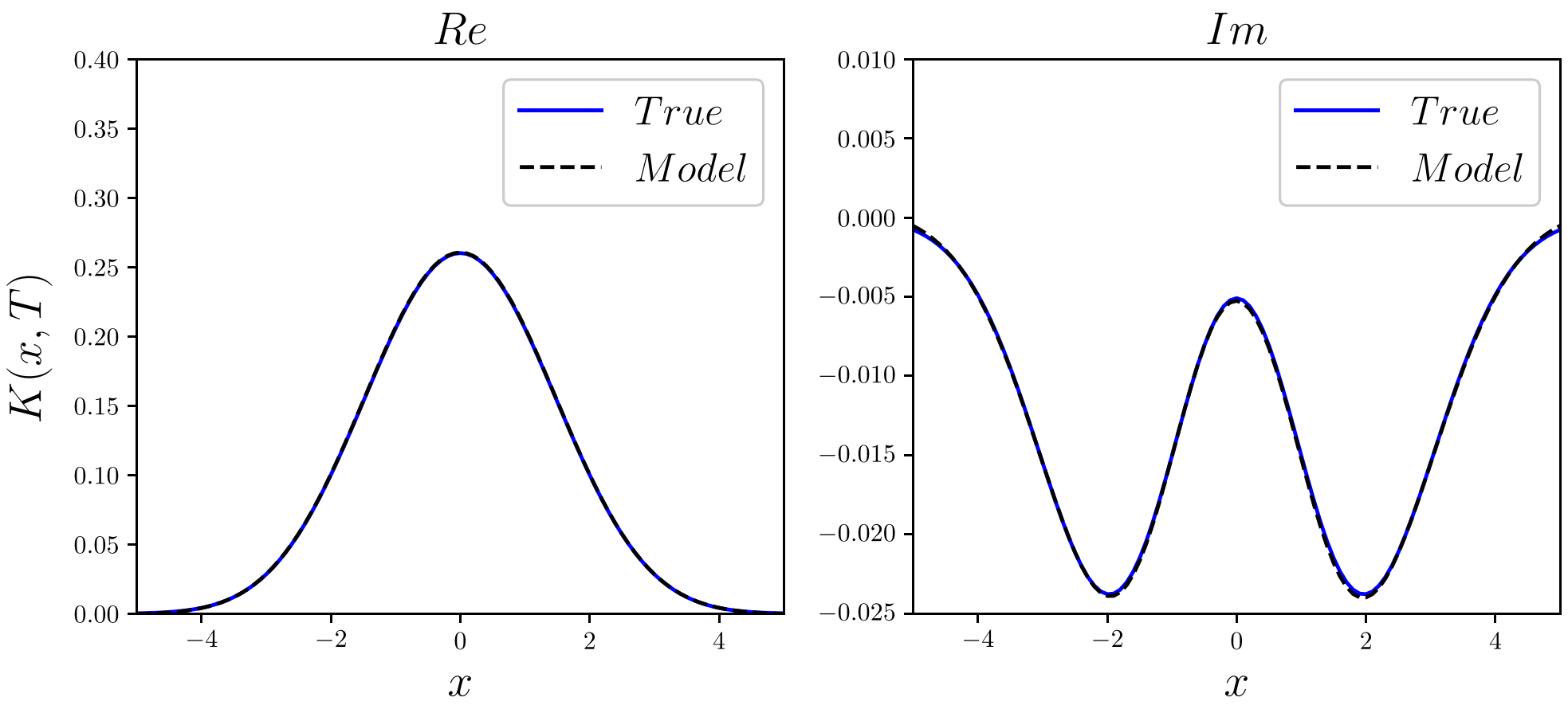

**Fig. 18.** Comparison of the true and estimated real and imaginary parts of the Euclidean propagator at $T = 1$.

Because of the very small uncertainty of the model, the approximated errors are not shown in the plot.

As expected, the neural network model is capable of approximating both the real and imaginary parts of the propagator; therefore, it can be used to describe the given path integral with complex potentials in the operating range that is set by the training process.

## 5. Approximation capabilities and time complexity of the neural network model

### 5.1. *Approximation capabilities and error bounds of the RBF model*

In this section, we will give more details on the approximating capabilities of the radial basis approximation given in Eq. (13), then discuss the time complexity and the possible advantages of the MLP model. The starting point of the model is the capability to approximate the functional given by:

$$\exp\Big(-\int_0^T d\tau\, V(x)\Big) \approx \sum_{k=1}^{M} a_k \exp\Big(-\int_0^T d\tau\, b_k(x-c_k)^2\Big), \tag{64}$$

where the main difficulty arises because of the integral in the exponents, which makes it difficult to say anything rigorous about the universal approximating capabilities of such a model. First, let us write both sides in a discretized form as:

$$\prod_{i=1}^{N} \exp\Big(-\Delta t\, V[x_i]\Big) \approx \sum_{k=1}^{K} a_k \prod_{i=1}^{N} \exp\Big(-\Delta t\, b_k(x_i-c_k)^2\Big) \tag{65}$$

where the $x$ coordinate is now discretized as $x_1, x_2, \ldots, x_N$, with a small finite time step $\Delta t$. Next, let us determine the behavior of the absolute difference between the true and approximated values, for which we will use the following notation for simplicity:

$$f_i = \exp\Big(-\Delta t\, V[x_i]\Big), \qquad g_{k,i} = a_k^{1/N} \exp\Big(-\Delta t\, b_k(x_i-c_k)^2\Big). \tag{66}$$

With this notation the error that we would like to express can be formulated as:

$$E = \left|\prod_{i=1}^{N} f_i - \sum_{k=1}^{K}\prod_{i=1}^{N} g_{k,i}\right|, \tag{67}$$

where we have used the fact that the $a_k$ factor in Eq. (65) can be put inside the product as $a_k^{1/N}$. Our task is to give an upper bound of this error and see how it behaves with $K$, $N$, $a_k$, $b_k$, and $c_k$. The first realization is that the RBF approximation of the full product in Eq. (65) is just the diagonal element of the more general case given by:

$$\prod_{i=1}^{N}\sum_{k=1}^{K}\alpha_k \exp\Big(-\beta_k(x_i-\gamma_k)^2\Big) = \sum_{\langle k\rangle\in K^N}\prod_{i=1}^{N}\alpha_{\langle k\rangle_i}\exp\Big(-\beta_{\langle k\rangle_i}(x_i-\gamma_{\langle k\rangle_i})^2\Big), \tag{68}$$

where $\langle k\rangle \in K^N$ means a sum over all of the $K^N$ combinations of the different parameters that appear when we expand the product of sums. The diagonal elements correspond to the case when $\alpha_{\langle k\rangle_i}$ is the same for all $i$, which in the case of the original RBF approximation means $\alpha_{\langle k\rangle_i} = a_k^{1/N}$ $(\forall i)$. This general approximation with $K^N$ terms corresponds to the case when we have one RBF model at each lattice site, which boils down to a simple function approximation problem for which the universal approximation properties of RBF networks hold. In the case of our "restricted" approximation with only diagonal terms, this is not that straightforward to see at this point. Next, let us express the error with inserting the general case as well and use the triangle inequality to get:

$$E = \left|\prod_{i=1}^{N} f_i - \prod_{i=1}^{N} G_i + \prod_{i=1}^{N} G_i - \sum_{k=1}^{K}\prod_{i=1}^{N} g_{k,i}\right| \le \underbrace{\left|\prod_{i=1}^{N} f_i - \prod_{i=1}^{N} G_i\right|}_{E_1} + \underbrace{\left|\prod_{i=1}^{N} G_i - \sum_{k=1}^{K}\prod_{i=1}^{N} g_{k,i}\right|}_{E_2} \tag{69}$$

where $f_i$ is the true function, $G_i = \sum_{k=1}^{K}\alpha_k e^{-\beta_k(x_i-\gamma_k)^2}$ represents the full, general RBF approximation at one lattice site, and $g_{k,i}$ corresponds to the diagonal terms of the full RBF expansion. The problem now is to estimate the upper bounds of the $E_1$ and $E_2$ errors, which can be done as follows.

First, let us assume that $x_i \in [-R, R]$, and the $f_i$ function is bounded as $\|f_i\|_\infty = C < \infty$, which is true for potentials that are bounded from below because of the $f_i = \exp(-V(x_i))$ exponential form. Starting with $E_1$, which represents the error of the full RBF approximation at each lattice site with $K^N$ terms, an upper bound can be given by using $|\prod_i a_i - \prod_i b_i| \le \sum_j |a_i - b_i| \prod_{i<j}|a_i| \prod_{i>j}|b_i|$, in which case we will arrive at:

$$E_1 \le \sum_{j=1}^{N} |f_i - G_i| \prod_{i<j}|f_i| \prod_{i>j}|G_i| \le N\epsilon L^{N-1}, \tag{70}$$

where $\epsilon$ is a small constant coming from the fact that the full RBF network at one lattice site can approximate any $f_i$ function with any desired accuracy $\|G_i - f_i\|_\infty \le \epsilon$, which is just the universal approximating theorem of RBF networks in function approximation problems, while $L^{N-1} = C^{N-1}(C+\epsilon)^{N-1}$ comes from the upper bound of the $\prod_{i<j}|f_i|\prod_{i>j}|G_i|$ terms. It is clear that at any fixed $N$ this error goes to zero when $\epsilon \to 0$, which means that with a sufficiently complex RBF network (more kernels, good distribution of centers, etc.) $E_1$ can be restricted to very small values.

However, the full error also depends on $E_2$, which gives the difference of the full RBF network defined at each lattice site and its diagonal parts. An upper bound of this error can be expressed as:

$$E_2 = \left|\prod_{i=1}^{N} G_i - \sum_{k=1}^{K}\prod_{i=1}^{N} g_{k,i}\right| \le \prod_{i=1}^{N}|G_i| + \sum_{k=1}^{K}\prod_{i=1}^{N}|g_{k,i}| \le (KA)^N + KA^N, \tag{71}$$

where the $(KA)^N$ factor comes from the fact that $\prod_i |G_i| = \prod_i \sum_k |\alpha_k| e^{-\beta_k (x_i-\gamma_k)^2} \leq \prod_i \sum_k |\alpha_k|$ for every $\beta_k > 0$, in which case $\prod_i |G_i| \leq (KA)^N$ in the case of $\alpha_k \leq A$ $(\forall k)$. The second term in turn comes from the fact that $\sum_k \prod_i |g_{k,i}| = \sum_k \prod_i |\alpha_k| e^{-\beta_k (x_i-\gamma_k)^2} \leq KA^N$, where the parameters of the diagonal terms of the full RBF expansion are defined as $\alpha_k = a_k^{1/N}$, $\beta_k = b_k$, and $\gamma_k = c_k$ from Eq. (65). According to this result, $E_2$ can be controlled by restricting $K$ and $A$ so that $KA < 1$, which means $\sum_k |a_k| < 1$ in the full RBF network applied to one lattice site. This constraint only limits the possible maximum that the approximation could reach using $K$ Gaussian kernels, which in general is not a problem because of the possible scaling of the path integral by a constant factor. Note that this is a worst-case scenario, and in practice the limitation is not that severe; however, in the case in which the network is struggling to learn a specific function, it is worth checking if it is because of this issue by applying different scaling on the approximable functions.

Using the obtained constraint for $E_1$ and $E_2$, the full error could be expressed as follows:

$$E = E_1 + E_2 \leq N\epsilon L^{N-1} + (KA)^N + KA^N, \tag{72}$$

where the first term is controlled by the accuracy of the full RBF network, while the second and third terms come from the difference between the full RBF network in Eq. (68) and the constrained network in Eq. (65).

There are a few things to address here. As stated before, the full RBF network, where one RBF network is applied at each lattice site, is able to approximate any "well behaved" (e.g. bounded on a finite interval) functions at a given $N$, owing to its universal approximation property. The situation with the constrained RBF network that corresponds to the diagonal of the full RBF is not that straightforward. From the corresponding error, it can be seen, however, that by choosing the parameters in a "good" way, i.e. $\sum_k |a_k| < 1$, it is possible to constrain the errors as $(\sum_k |a_k|))^N + \sum_k (|a_k|^N)$. Apart from this worst-case scenario, it is still not trivial how many kernels we need or what the range of $a_k$, $c_k$, and $b_k$ should be to be able to approximate a specific functional form. This is a general problem in any neural network, and the answer is always given by testing out different configurations with different numbers of neurons, different scales, different inputs, etc. Nevertheless, it is very important to give at least a crude approximation of the capabilities of the model. In this case the model is expected to work for a large set of bounded potentials defined on a finite range $x$. Because of the exponential damping, the effective $x$ can always be set to be between $x \in [x_{\min}, x_{\max}]$, and the path integral can be scaled so that the diagonal terms dominate in the full RBF for large enough $N$.

It was shown before that the method is able to approximate $e^{-V(x)}$ in the case of a simple quartic potential. Now let us test the model on something more complex and define the potential as:

$$V(x) = 1 + \sin(2x) + 0.2x^2 + e^{-10(x-2)^2}. \tag{73}$$

In Fig. 19 we show the potential and also $e^{-V(x)}$, which could give us an estimate of the coordinate interval, which we need to generate the training data. From $e^{-V(x)}$ it is seen that $x \in [-6, 6]$ is probably sufficient to generate the dominant paths. The given potential now has multiple local minima and does not have discrete $Z_2$ symmetry as in the quartic case; however, it is still continuous and bounded from below, which are the key properties for the RBF approximation to work.

We have set the Euclidean time to $T = 0.5$, the range to $x \in [-6, 6]$, and the number of neurons in the hidden layer to $M = 100$, with the same number of inputs ($N = 7$) as before with centers $c_i = [-5, -3, -1, 0, 1, 3, 5]$. In Fig. 20 the training process can be followed using 50 000 training and 10 000 validation samples, while in Fig. 21 the comparison of the true and approximated values for 100 test samples is shown together with the (normalized) distribution of the differences calculated for 15 000 pieces of test data.

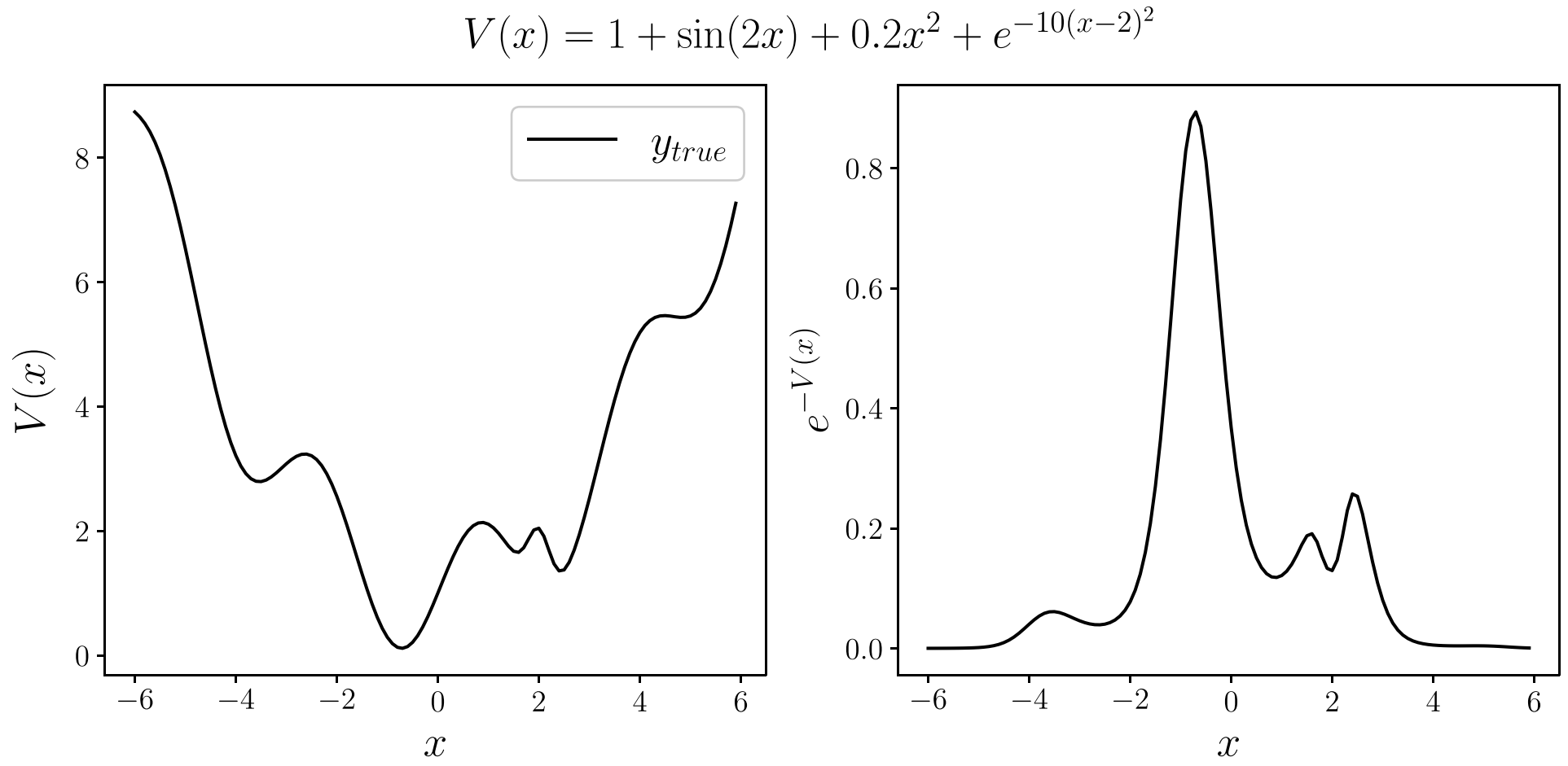

**Fig. 19.** The potential $V(x)$ (left) and $e^{-V(x)}$ (right) that will be used to test the capabilities of the RBF approximation.

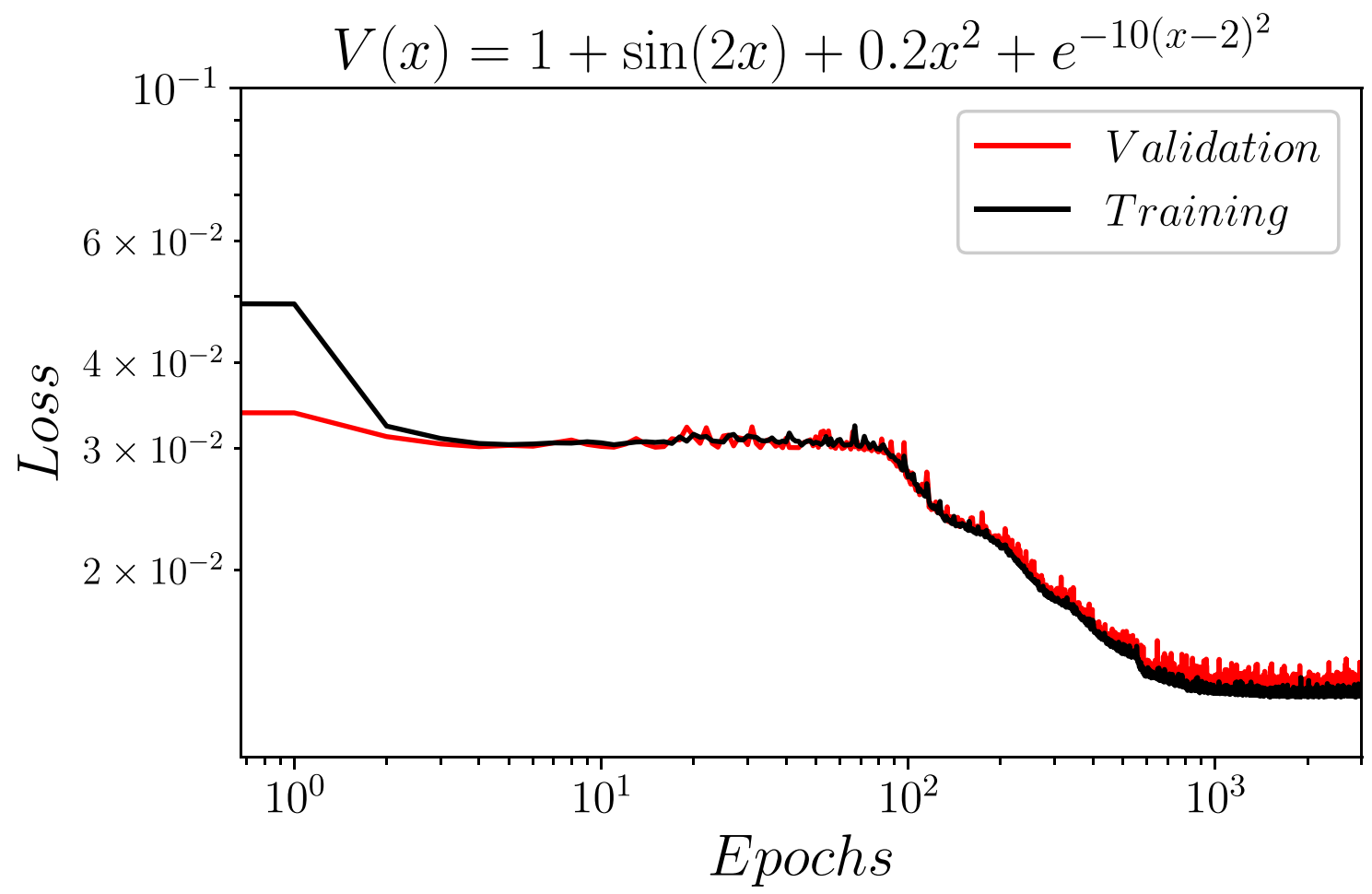

**Fig. 20.** Evolution of the training and validation losses for the $V(x) = 1 + \sin(2x) + 0.2x^2 + e^{-10(x-2)^2}$ potential using $M = 100$ neurons in the hidden layer. The Euclidean time is set to $T = 0.5$.

The results show that the network is able to give the same order of magnitude approximation to this more complex potential as in the case for the simple quartic potential case. In the next subsection we will examine the time complexity of the model in the context of the bound state wave function calculation that has been carried out in Section 4.

### 5.2. *Time complexity of the bound state calculations*

One of the main advantages of the neural network model is the possibility to "instantly" obtain the propagators at different times and for different parameters after training the model. It is, however, not straightforward to identify how the time of the training procedure scales with the number of inputs, the number of neurons, or the complexity of the potential. Here, we will briefly address this issue and compare the time that it takes to obtain the propagators with the RBF model to the time that it takes to get the same result with solving the bound state problem directly through discretizing the Schrödinger equation. All the calculations in this paper have been carried out on a standard LG Gram notebook equipped with an Intel Core i5 processor and 16 GB of memory, with no direct GPU support or parallel computing.

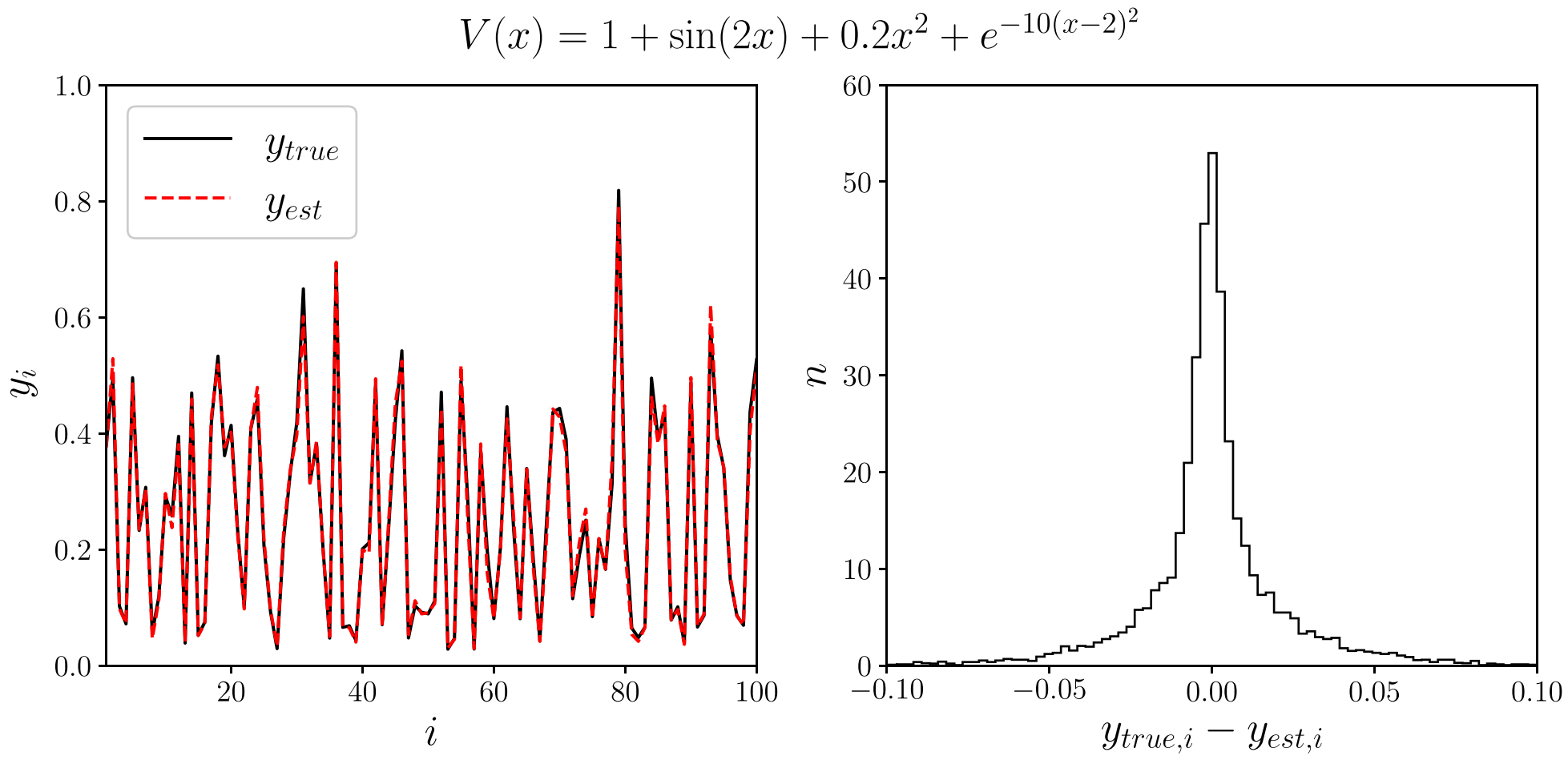

**Fig. 21.** The comparison of the true and estimated values for 100 test samples (left), and the normalized histogram of the differences ($y_{\text{true},i} - y_{\text{est},i}$) (right) using 15 000 test samples.

Obtaining the propagator from the neural network model has three main steps: (1) generating the training samples, (2) training the network, and (3) calculating the propagator. From these, (1) takes only a few seconds to generate 100 000 samples, while (3) is below a second; thus the only term that we need to consider is (2), which could only be determined by trying out different configurations with different sets of training samples.

We will use the system described in the last example in Section 4.1 as a benchmark, in which case the potential is set to $V(x) = 0.05x^4 - x^2$, the Euclidean time is set to be $T \in [2.5, 3]$, and the interval of the boundary values is given between $x_i, x_f \in [-6, 6]$. The number of inputs of the network is fixed to 7 with centers $c_i = [-5, -3, -1, 0, 1, 3, 5]$. The training/validation/test samples (paths) are generated in the range of $u(\tau) \in [\max(x - 3, -6), \min(x + 3, 6)]$ for every $\tau \in [0, T]$ using the same PCHIP method as before. In our tests, we will vary the number of trainingbatchv/validation samples and the number of neurons in the hidden layer ($M$) and analyze the scaling of the training time at a fixed number of epochs and at a fixed accuracy. The training of the model is carried out with the Adaptive Moment Estimation (ADAM) algorithm, which is a widely used gradient based optimization algorithm that is very robust and fast even with large datasets. In addition we have used batch training with a fixed batch size $N_{\text{batch}} = 500$ in every case.

For the first test we have fixed the number of epochs to $N_{\text{epoch}} = 500$ and varied the number of training/validation samples ($N = N_{\text{train}} + N_{\text{valid}}$, with $N_{\text{valid}} = 0.2N$) and the number of neurons $M$ to see how the time of training scales with $N$ and $M$. The results can be seen in Fig. 22 for $M = 10, 50, 1000$, and $N = 5000, 25\,000, 50\,000, 75\,000, 1000\,000$.

The result shows the expected linear behaviour with an increasing number of samples, and with an increasing number of neurons. The $N_{\text{epoch}} = 500$ stopping condition, however, does not say anything about the accuracy of the model; thus these results do not correspond to a valid training procedure, and only give us some insight into the behaviour of the traning method, not the model itself.

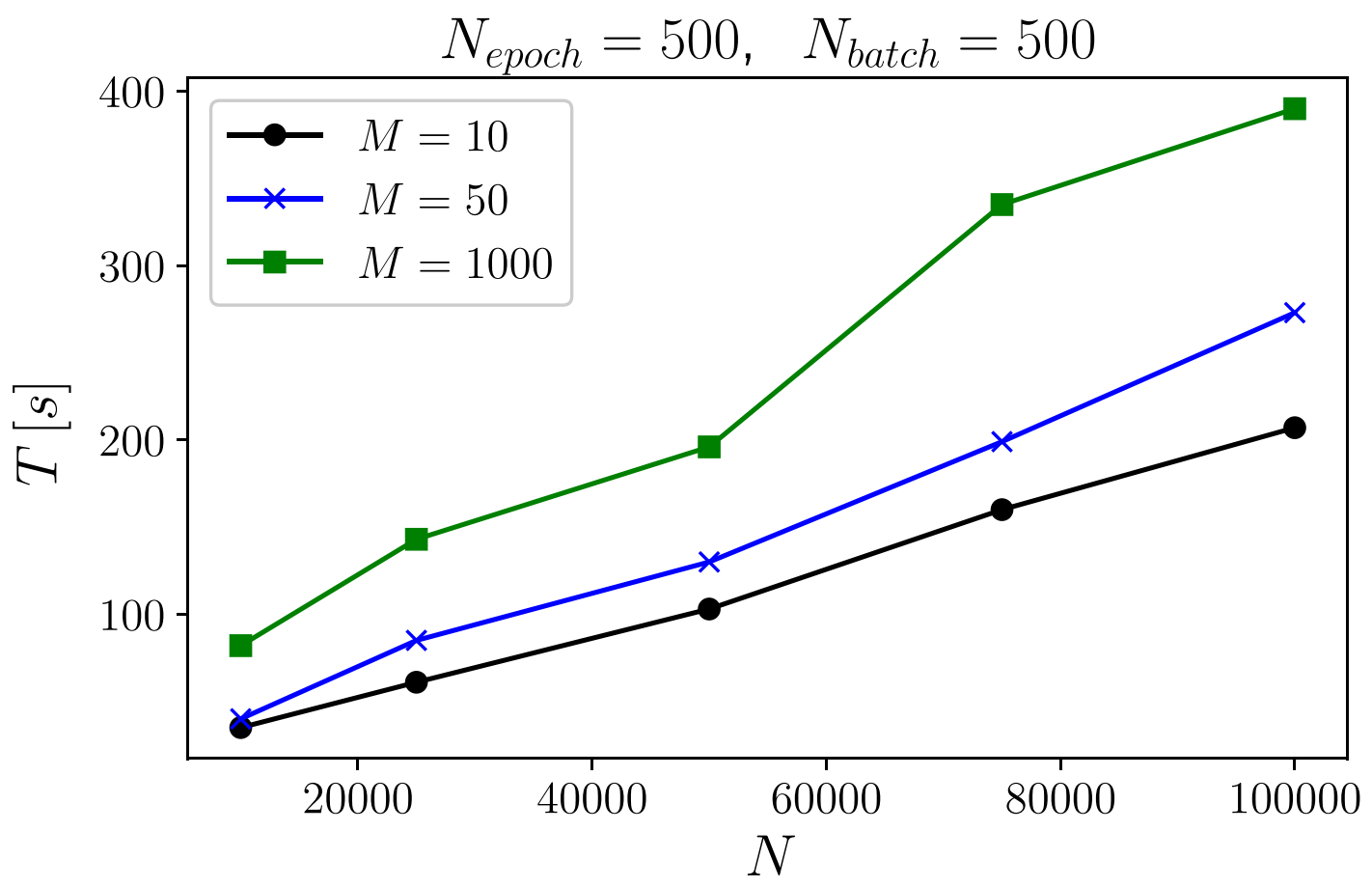

**Fig. 22.** Dependence of the neural network training time on the number of training samples, and on the number of neurons in the hidden layer.

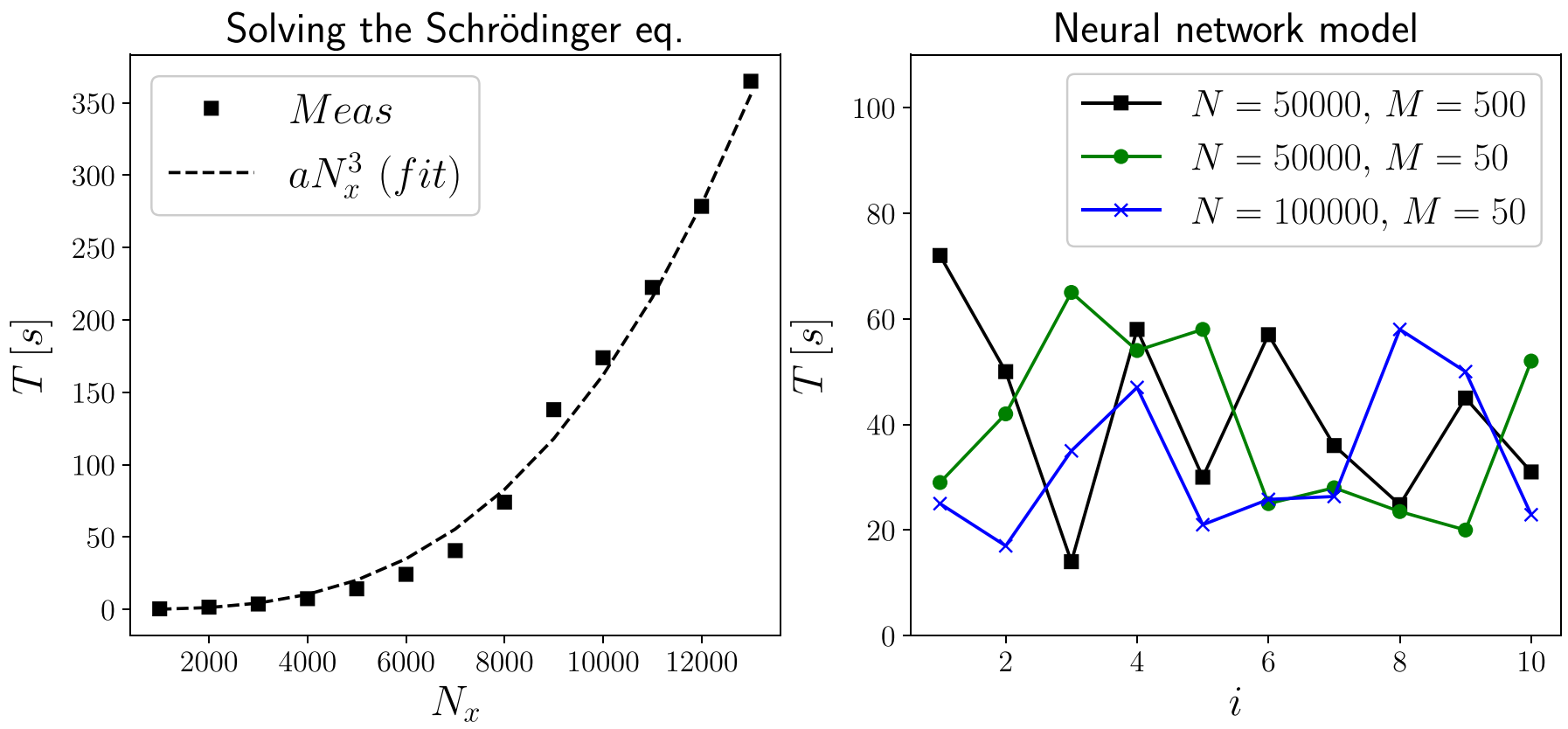

**Fig. 23.** Time for obtaining the bound state wave functions from solving the discretized Schrödinger equation (left) using $N_x$ lattice points, and the time to train the corresponding neural network model (right) for 10 different initializations, with different training samples, using three cases with different $(N, M)$.

The time that it takes to train the model up to predetermined accuracy depends on many factors, e.g. complexity of the potential, complexity of the network, batch size, number of training samples, etc. In general it is not possible to make an estimate without doing some test training on different datasets. For example, even though we have seen that, in the case of increasing $N$, the time that it takes to reach a certain number of epochs is a linearly increasing function of $N$, this will not necessarily be the case when we want to reach the same accuracy using two different numbers of training sets. It is possible that, because of the larger training and validation sets, the convergence will be better; thus it takes a smaller number of steps to reach the same accuracy. This also applies to the number of neurons $M$, or the number of epochs, and so on. In the right-hand plot in Fig. 23, the results for the elapsed time during training that it takes to reach the same accuracy as in Fig. 12 are shown using 10 different initializations and training/validation samples for three cases: $(N, M) = (50\,000, 500)$, $(N, M) = (50\,000, 50)$, and $(N, M) = (100\,000, 50)$.

From these results it is evident that the previously observed scaling is not as dominant as the convergence properties change by changing the parameters and/or the number of samples that are used in training. The elapsed time fluctuates around 40 s in each case, which is a very short time for the training of a neural network. It has to be noted that we could even reach lower errors by not stopping the training at this accuracy; however, as the convergence slows down after this region, it could take a very long time to do so. As we do not need better accuracy to get the results shown in Figs. 14 and 15, this stopping condition is sufficient.

On the left-hand side the time of the bound state calculation is also shown by solving the discretized Schrödinger equation with $N_x$ lattice points, using QR decomposition to solve the corresponding eigenvalue problem that generally has an $\mathcal{O}(N_x^3)$ time complexity. The measurements show a clear cubic dependence on the $N_x$ lattice points, which blows up after around $N_x \approx 10\,000$. For smooth, not too oscillating potentials, as in the case of $V(x)$ in this example, only a few thousand lattice points are sufficient to get the bound state wave functions with good accuracy; however, in more complex cases (or even for complex-valued potentials), this may not be the case, and more $N_x$ could be needed.

From these results it is evident that for this exact problem the neural network model gives very promising results. The main advantage lies in the fact that after one training, we could extract not just the wave function itself but other observables that can be calculated from the time dependence of the propagator as well. This can be done "instantly" without retraining the model, even for different masses in the operating range of the model. It is, however, not straightforward to see how the time complexity scales with the properties of the different potentials. For smooth, bounded potentials in the Euclidean time ranges that were shown in this paper, the training is very fast and can be done on average in under a minute. However, for larger operating ranges, i.e. larger Euclidean times, very small masses, or for more exotic potentials with e.g. discontinuities defined on larger intervals, etc., this training could take a much longer time to get a satisfying accuracy. It is thus a very important modeling step to set up a corresponding network for a specific system.

## 6. Conclusions

In this paper we have shown a method on how to approximate Euclidean path integrals that have a general family of real and/or imaginary potential terms by a feedforward neural network architecture that represents a quadratic expansion of the corresponding interaction terms. The training samples are generated by using piecewise Hermite polynomial interpolation between randomly generated points, giving well behaved, continuous, and easily adjustable functions that can be used to obtain good generalization of the model. This generalization is a very important aspect of the model, as the path integral in essence requires the infinite sum over all possible paths that are allowed by symmetries and the boundary conditions. Extracting the trained parameters of the neural network allows us to approximate the original path integral by the finite sum of shifted harmonic oscillators (and a term for the free propagator) that have analytical solutions; therefore, it allows us to calculate observables in a closed form. The method has been tested for a double-well-type potential, where we have compared the bound state wave functions that are extracted from the trace-normalized propagator at large Euclidean times with the numerical solution of the Schrödinger equation, giving a very good match within a few per-centage relative error. The neural network model has also been tested for complex potentials, where we have examined the problem of the harmonic oscillator with complex frequencies and trained an extended neural network construction with complexified coefficients that was able to approximate the real and imaginary parts of the propagators as a multiple input–multiple output system. The results for the real and imaginary parts of the Euclidean propagators at a specific Euclidean time have been compared to the analytical solution, giving a very good match between the approximation and the true solution within one percentage of relative error.

Owing to the general nature of the neural network model, it would be possible to extend the method to describe path integrals in quantum field theories as well, where the field configurations will be taking the role of the paths that have been used in nonrelativistic quantum mechanics. In the case of higher dimensions, the training procedure of the network for such systems is not that trivial, as in the case of 1D quantum mechanics. In Refs. [59] and [60], a slightly different approach to the same method is applied to the lattice formulation of complex and real scalar field theories with quartic interactions, where the system is approximated by one RBF network on each lattice site. That description is an extension of the model shown here and is more suitable to describe field theories on the lattice. In Ref. [59], it was also demonstrated that employing the lattice formulation of the RBF expansion for complex scalar fields at finite densities effectively resolved the sign problem in a numerically efficient manner, while also providing an accurate description of phase transitions at finite chemical potentials and the Silver Blaze phenomenon.

There is a possibility that the method could help address the long-time numerical problem of quantum chromodynamics at finite densities, which would be a great step towards understanding many related phenomena regarding phase transitions and the quark–gluon plasma, the nuclear equation of state of neutron stars, modification of particle masses and widths at finite densities, etc. The applicability of the method to such problems is yet to be seen and will be the topic of future work.

## Funding

This work was supported by the Korea National Research Foundation under Grant No. 2023R1A2C300302311 and 2023K2A9A1A0609492411. The author was supported by the Hungarian OTKA fund K138277. Open Access funding: SCOAP$^3$.

Funded by SCOAP[3]